\documentclass{article}

\usepackage{arxiv}

\usepackage[utf8]{inputenc} 
\usepackage[T1]{fontenc}    
\usepackage{doi}            
\usepackage{graphicx}%
\usepackage{multirow}%
\usepackage{amsmath,amssymb,amsfonts}%
\usepackage{amsthm}%
\usepackage{mathrsfs}%
\usepackage[title]{appendix}%
\usepackage{xcolor}%
\usepackage{textcomp}%
\usepackage{booktabs}%
\usepackage{algorithm}%
\usepackage{algorithmicx}%
\usepackage{algpseudocode}%
\usepackage{listings}%
\usepackage{framed}
\usepackage{tabularx}
\usepackage{pifont}
\usepackage{latexsym}
\usepackage{rotating}
\usepackage{tablefootnote}
\usepackage{threeparttable}
\usepackage{array}
\usepackage{soul}
\usepackage{lscape}
\usepackage{multicol}
\usepackage{afterpage}
\usepackage{dblfloatfix} 
\usepackage{afterpage}
\usepackage{changepage}
\usepackage{url}
\usepackage[numbers]{natbib}

\title{Foundation Models in Medical Imaging -\\
A Review and Outlook}

\author{
\textbf{Vivien van Veldhuizen}\textsuperscript{1,2} \quad
\textbf{Vanessa Botha}\textsuperscript{1,3} \quad
\textbf{Chunyao Lu}\textsuperscript{1,2} \quad
\textbf{Melis Erdal Cesur}\textsuperscript{1} \\
\textbf{Kevin Groot Lipman}\textsuperscript{1} \quad
\textbf{Edwin D. de Jong}\textsuperscript{4} \quad
\textbf{Hugo Horlings}\textsuperscript{1} \quad
\textbf{Cl\'arisa Sanchez}\textsuperscript{5} \\
\textbf{Cees Snoek}\textsuperscript{5} \quad
\textbf{Lodewyk Wessels}\textsuperscript{1,3,6} \quad
\textbf{Ritse Mann}\textsuperscript{1,2} \quad 
\textbf{Eric Marcus}\textsuperscript{1,5}\thanks{These authors contributed equally to this work.} \quad
\textbf{Jonas Teuwen}\textsuperscript{1,2,5}\footnotemark[1]
\vspace{0.1cm}\\
\textsuperscript{1}Netherlands Cancer Institute \quad
\textsuperscript{2}Radboud University Medical Center \\
\textsuperscript{3}Delft University of Technology \quad
\textsuperscript{4}Kaiko.ai \quad 
\textsuperscript{5}University of Amsterdam \quad
\textsuperscript{6} Oncode Institute \\
{\tt\small j.teuwen@nki.nl, v.v.veldhuizen@nki.nl}
}

\newcommand{\cmark}{\ding{51}}%
\newcommand{\xmark}{\ding{55}}%

\begin{document}
\maketitle

\begin{abstract}
Foundation models (FMs) are changing the way medical images are analyzed by learning from large collections of unlabeled data. Instead of relying on manually annotated examples, FMs are pre-trained to learn general-purpose visual features that can later be adapted to specific clinical tasks with little additional supervision. In this review, we examine how FMs are being developed and applied in pathology, radiology, and ophthalmology, drawing on evidence from over 150 studies. We explain the core components of FM pipelines, including model architectures, self-supervised learning methods, and strategies for downstream adaptation. We also review how FMs are being used in each imaging domain and compare design choices across applications. Finally, we discuss key challenges and open questions to guide future research.
\end{abstract}

\section{Introduction}

Medical image analysis plays a crucial role in modern healthcare, with techniques such as X-ray, magnetic resonance imaging, and microscopic tissue analysis providing essential insights for diagnosis, treatment planning, and disease monitoring. A persistent challenge in this field is the scarcity of labeled data, which is expensive and time-consuming to obtain due to privacy concerns, expert annotation requirements, and the complexity of acquiring high-quality labels. Traditional supervised learning methods, which rely heavily on large annotated datasets, often struggle under these constraints.

Foundation Models (FMs), a recent development within Artificial Intelligence (AI), offer a promising solution. Pre-trained on large, diverse, unlabeled datasets, FMs learn rich, general representations that capture broad patterns and features. Fine-tuning these models on smaller, task-specific labeled datasets reduces dependency on extensive annotations while often improving performance compared to traditional approaches.

Despite their potential, the application of FMs in medical image analysis is still emerging. This review focuses on vision-based FMs for medical imaging across three primary domains: pathology, radiology, and ophthalmology. By systematically analyzing over 150 studies, we provide a comprehensive overview of the current state of research and the various models developed and employed across these domains.

\subsection{Overview}
Section \ref{section:background} introduces foundation models and their typical deployment pipeline, covering three core concepts: large-scale architectures, self-supervised learning, and adaptation to downstream tasks. Sections \ref{section:pathology}, \ref{section:radiology}, and \ref{section:ophthalmology} review FMs in pathology, radiology, and ophthalmology, respectively, including domain introductions and tables summarizing landmark works. Section \ref{section:challenges} discusses key challenges and open questions to guide future research.

Our contributions are summarized as follows:
\begin{itemize}
    \item \textbf{Foundation Model Background}: In-depth explanation of FMs and their pipeline divided into three main building blocks.
    \item \textbf{Comprehensive Review}: Examination of over 150 studies with summary tables for quick reference.
    \item \textbf{Modality-Specific Organization}: Review organized by the main imaging domains (pathology, radiology, ophthalmology) for detailed insights.
    \item \textbf{Identification of Challenges}: Highlighting key challenges and open questions to guide future research.
\end{itemize}

\subsection{Relation to Other Works}
Several recent reviews have addressed foundation models in healthcare. Some take a broad perspective, covering applications beyond imaging, such as natural language processing for electronic health records, drug discovery, medical robotics, and education \citep{zhang2023challenges, qiu2023large, khan2024comprehensive, zhang2024data}. Others concentrate primarily on large language models, without substantial coverage of vision-based foundation models \citep{liu2024large, yuan2023large, zhou2023survey, he2023survey, thirunavukarasu2023large}. Within medical imaging, \citet{azad2023foundational} provide a general overview but do not analyze individual studies in detail, while \citet{paschali2025foundation} focus specifically on radiology and emphasize overarching concepts rather than surveying the breadth of implementations.

Our review differs in two key ways. First, we focus exclusively on vision-based foundation models for medical image analysis, excluding language-only models and broader healthcare applications. Second, we provide both a wide coverage and a structured analysis: over 150 studies are systematically reviewed, organized by imaging domain (pathology, radiology, and ophthalmology), and presented with summary tables and detailed discussion. This structure allows us to place individual contributions in context, highlight methodological trends and domain-specific differences, and identify open challenges. In this way, our work complements prior reviews by offering a comprehensive and modality-specific resource for researchers in medical imaging.

\subsection{Methodology}
For a comprehensive review of foundation models in medical image analysis, we performed a systematic search of relevant literature across multiple databases. Our search included PubMed, ArXiv, MedArxiv, and Google Scholar. The query used was:

\begin{quote}
\texttt{"foundation model" OR "foundational model" OR "self-supervised model"OR "self-supervised" OR "self-supervised deep" OR "SSL model" OR "large language model" OR "large-language model" OR "self supervised learning"}
\end{quote}

The search was conducted until January 2025. Inclusion criteria required that studies address the application of foundation models in medical imaging or employ self-supervised learning techniques on a large scale. Exclusion criteria removed studies that did not explicitly apply to the medical domain, such as those related to biological and biomedical fields, or were focused on models that exclusively processed non-imaging modalities, such as language, video, or genomic sequences. 

Titles and abstracts were initially screened by two independent reviewers to identify potentially relevant articles. Full-text reviews were then performed to confirm the suitability of these articles for inclusion in the review.

\section{Background} \label{section:background}

    \paragraph*{What is a Foundation Model?}
    The term \textit{Foundation Model} was popularized by the Stanford Institute for Human-Centered Artificial Intelligence, defining them as ``any model that is trained on broad data (generally using self-supervision at scale) that can be adapted (e.g., fine-tuned) to a wide range of downstream tasks'' \citep{bommasani_opportunities_2022}.
    
    \paragraph{Scaling Laws}
    The development of foundation models has been significantly influenced by the realization that larger datasets lead to improved model performance. This understanding is supported by the concept of scaling laws, which describe predictable improvements in performance as model size, dataset size, and computational resources are increased \citep{kaplan2020scaling}. Specifically, as datasets grow larger and model parameters and computational resources are scaled accordingly, models generally exhibit enhanced performance \citep{achiam2023gpt}. In the general domain, FM network architectures can have trainable parameters in the range of billions, and dataset sizes are also substantial, ranging from tens of thousands to hundreds of millions of images. For example, OpenAI’s CLIP model was trained on 400 million image-text pairs \citep{radford2021learning}. 
    Given the limitations of obtaining large labeled datasets, especially in specialized fields like medical imaging, the focus has shifted toward utilizing self-supervised learning (SSL) methods to leverage large amounts of unlabeled data, thus leading to the current foundation models paradigm.
    
    \paragraph{Emergent Properties}
    The scale of foundation models not only contributes to their generalizability but can also lead to the model exhibiting novel behaviors and insights that smaller models might not demonstrate. Such emergent abilities include improved understanding, generation, and contextualization of data, which can significantly enhance model performance across various tasks \citep{awais_foundational_2023}.
    
    The initial success of Large Language Models (LLMs), such as the GPT series from OpenAI \citep{brown_language_2020} and BERT from Google \citep{devlin_bert_2019}, marked the start of foundation models. Building on their success in natural language processing, the concept expanded into computer vision with the development of \textbf{\textit{Vision Foundation Models (VFMs)}} and \textbf{\textit{Vision-Language Foundation Models (VLFMs)}}, which integrate both visual and textual data. 

\paragraph*{Why Foundation Models for Medical Imaging?}
Traditionally, deep learning models for image analysis are trained using supervised learning, which requires large amounts of labeled data. In medical imaging, such annotated datasets are often scarce due to the time, cost, and expertise required for labeling, which severely limits the performance of deep learning models. Some methods try to overcome this limitation by pre-training on large labeled natural image datasets, such as ImageNet, which are more readily available. However, because natural images differ greatly from medical images in texture, structure, and context, these models often struggle to generalize across medical specialties and imaging modalities.

Foundation Models overcome this limitation by learning from large, diverse, \textit{unlabeled} datasets using self-supervised learning. Instead of being trained for a single task, they learn general representations that capture broad patterns and features in the data. These rich representations can then be fine-tuned for a variety of medical imaging tasks—such as classification, segmentation, detection, and diagnosis—using relatively few labeled examples. This approach reduces the reliance on annotated datasets while improving generalization across imaging modalities and clinical domains.

\subsection{FM Core Technical Concepts} \label{section:fm_building_blocks}

    \begin{figure}
        \centering
        \includegraphics[width=0.9\linewidth]{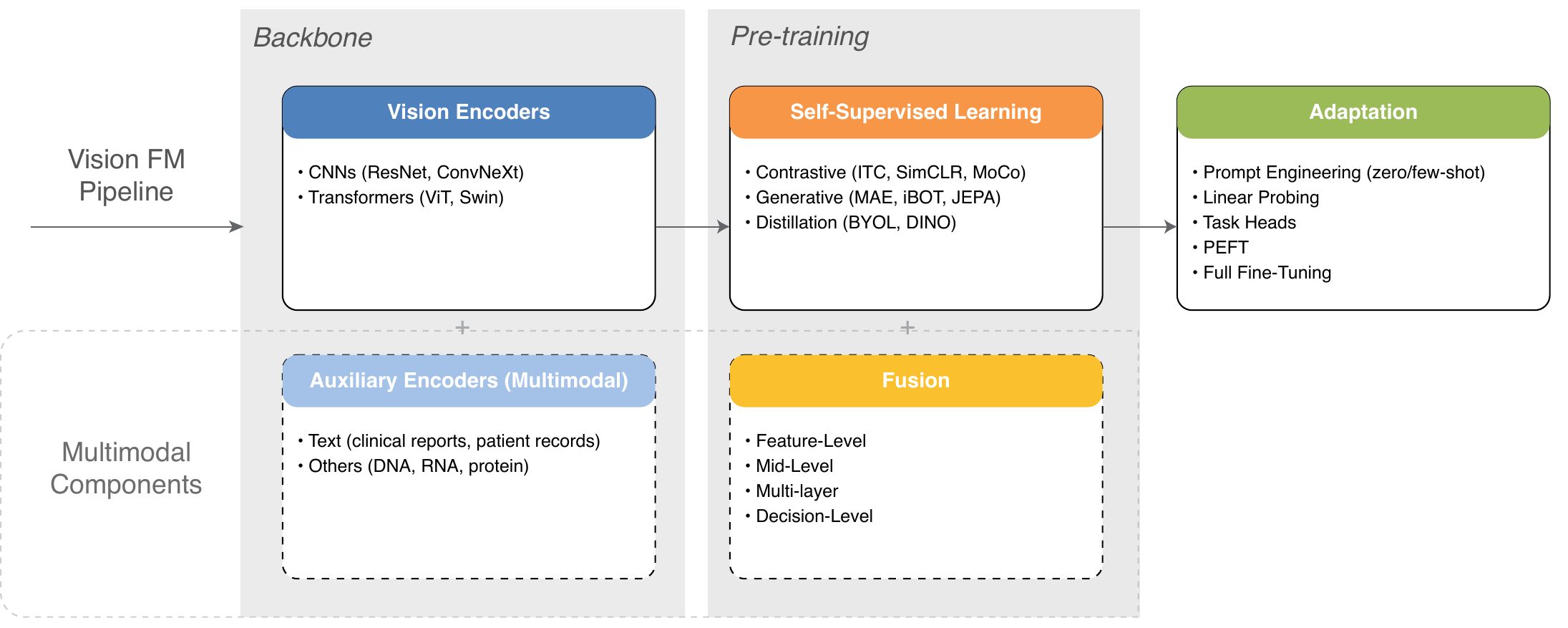}
        \caption{Overview of core technical concepts behind foundation models for medical image analysis. Dotted lines represent optional components. The backbone architecture consists of vision encoders processing imaging data, while auxiliary encoders incorporate complementary modalities in multimodal settings, such as clinical notes, patient records, or molecular data. Self-supervised learning methods, such as contrastive, generative, and distillation approaches, are used to pretrain the model and learn rich representations from the data. In multimodal scenarios, representations from different encoders can be fused before adaptation to downstream tasks. Adaptation strategies range from lightweight, zero/few-shot methods to heavier approaches such as full fine-tuning.}
        \label{fig:overview}
    \end{figure}
    
Foundation models are characterized by three core concepts that set them apart from traditional deep learning methods. At the foundation of these models' capabilities are: \textbf{1.} \textbf{\textit{large-scale pretraining}} on extensive datasets, enabled by \textbf{2.} \textbf{\textit{self-supervised learning}} objectives that extract meaningful patterns without manual annotation. Through this process, the models learn rich, generalizable representations in the form of \textit{embeddings}: high-dimensional vectors that capture clinically relevant patterns and relationships in the data. Unlike traditional deep learning models designed for specific tasks, foundation models organize knowledge in these versatile embeddings, which can then be \textbf{3.} \textbf{\textit{efficiently adapted}} to various downstream clinical applications through multiple approaches (see Figure \ref{fig:overview}). Below, we examine these concepts in detail.
    
\subsubsection{Large-Scale Pre-Training}
The purpose of large-scale encoders in foundation models is to extract meaningful patterns from large datasets and convert raw data into generalizable embeddings. Depending on the data type, different architectures are used: transformer-based models are typically employed for textual data, while visual data are processed using either \textit{Convolutional Neural Networks (CNNs)} or \textit{Vision Transformers (ViTs)}. In this review, we focus on the latter two families, which form the backbone of vision-based foundation models. 
    
    \paragraph{Convolutional Neural Networks}

    CNNs have been fundamental in computer vision and are known for their ability to extract hierarchical features from image data. They assume that nearby pixels are related (locality) and that different parts of an image are processed similarly (weight sharing). These inductive biases help CNNs learn effectively from limited training data, making them particularly efficient for tasks where local patterns, such as edges and textures, are crucial. ResNet \citep{resnet}, for example, is a popular choice for CNN-based foundation models due to its depth and residual connections, which help maintain the flow of information in deeper architectures. ConvNeXt \citep{liu2022convnet} is another modern adaptation of CNNs that incorporates some elements of transformer architectures to improve performance while retaining the beneficial inductive biases of CNNs.

    \paragraph{Transformers}
    The transformer architecture was originally developed for natural language processing \citep{vaswani2017attention} and later successfully adapted to visual data through ViTss \citep{dosovitskiy2021image}. Rather than using sliding convolutional filters, as CNNs do, ViTs split an image into fixed-size patches (for example 16×16 pixels) and treat each patch as an individual element in a sequence.
    These patches are first converted into feature vectors through a linear projection. Crucially, since this process loses spatial information, learned positional encodings are added to preserve where each patch was located in the original image. 
    The architecture's core innovation lies in its self-attention mechanism, which allows every patch to dynamically interact with every other patch, determining how much attention to pay to different parts of the image based on their relevance.
    This global attention mechanism allows the model to capture both local patterns and long-range dependencies equally effectively.
    
    Most implementations of foundation models in medical imaging use a simplified variant that focuses exclusively on building comprehensive image representations. These encoder models maintain the original architecture's ability to process all image patches simultaneously through bidirectional attention, where information flows freely between all patches. This differs from the more complex original design which included separate components for analysis (encoder) and generation (decoder), a distinction that is more relevant to tasks like machine translation than visual representation learning.   
    
    \paragraph{Choice of Backbone}
    Despite their differences, CNNs and transformers can be powerful backbone architectures for foundation models in various domains. CNNs leverage inductive biases like locality and weight sharing, allowing them to effectively extract spatial features and perform well with limited data. These biases make CNNs efficient for tasks where local patterns (e.g., edges, textures) are crucial, but they can limit the model's ability to capture long-range dependencies, which are often important in large-scale or complex datasets.
    In contrast, ViTs have minimal inductive biases, enabling them to model global dependencies across images. This flexibility allows ViTs to excel in large-scale pretraining, particularly with self-supervised learning, although it also requires significantly larger training datasets to achieve optimal performance \citep{goldblum2024battle}. As a result, ViTs are more commonly used as backbones for foundation models due to their scalability. However, CNNs remain effective in scenarios with limited data and are also integrated into hybrid architectures, such as Pyramid Vision Transformers (PvT), Swin Transformers, and Convolutional Vision Transformers (CvT) \citep{wang2021pyramid, liu2021swin, wu2021cvt}, which combine the strengths of CNNs and transformers to enhance data efficiency and multiscale feature representation.
    
    \begin{figure*}
        \centering
        \includegraphics[width=0.8\textwidth]{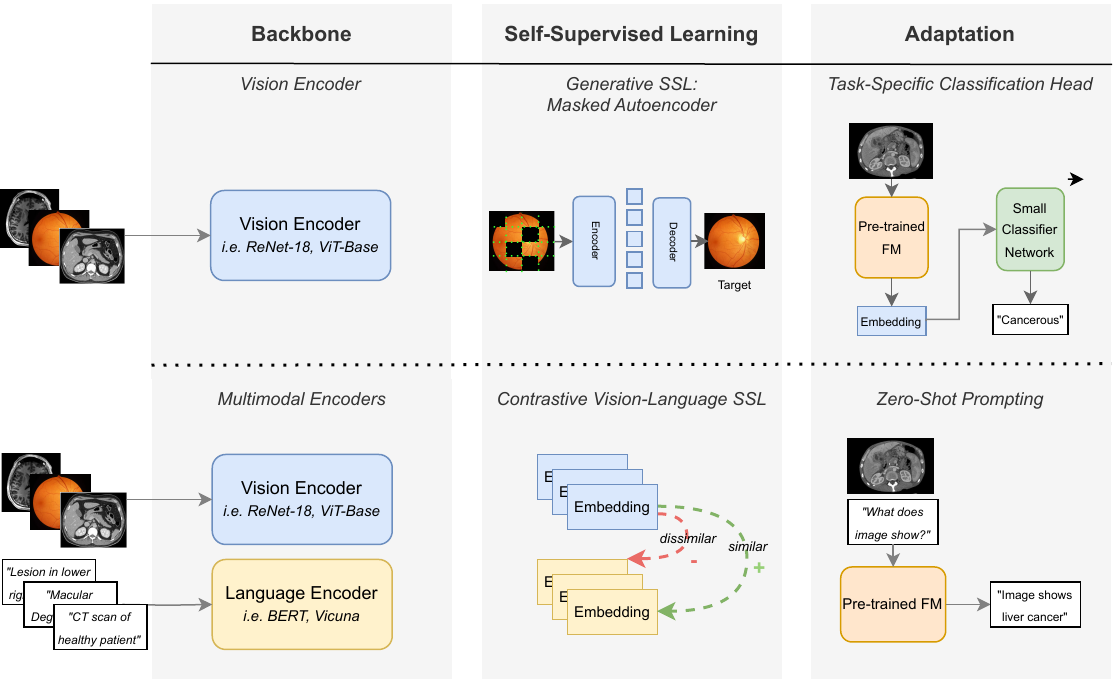}
        \caption{Two simplified examples of how different architectures, SSL techniques, and adaptation methods can be combined in foundation models for medical image analysis. \textit{Top row}: vision encoder backbone is pre-trained using masked autoencoding, in which it learns to reconstruct missing parts of the training images. The model learns to encode images in a meaningful way. After pre-training, a small labeled training dataset is encoded by the VFM, and these embeddings serve as input for training a small tumor classification head. \textit{Bottom row:} Both a vision encoder and text encoder are trained. In the pre-training phase, the VLFM learns to group similar images and text through a contrastive SSL objective. After pre-training, the model is prompted in a zero-shot way; it is given a new image with a corresponding question and must predict the answer without any additional fine-tuning.}
        \label{fig:enter-label}
    \end{figure*}    
    
\subsubsection{Self-Supervised Learning}
    One of the key components of foundation models is the use of self-supervised learning (SSL). Since the large amounts of data that FMs require don’t come with labels, SSL is essential for extracting patterns from unlabeled data. Unlike supervised learning, which relies on human-provided labels, SSL derives the supervision signal directly from the data, using its inherent structures or relationships.  At the core of SSL are pretext tasks, auxiliary tasks designed to provide supervision signals during training.
    By solving these pretext tasks, the model learns to extract useful features or representations from the input data, which can then be applied to downstream tasks.

    SSL approaches can be broadly categorized into discriminative and generative methods. Discriminative SSL distinguishes between different samples or representations, while generative SSL aims to predict missing information from the input. 

    \paragraph{Discriminative Self-Supervised Learning}

    Discriminative SSL methods can be further divided based on their underlying techniques. One of the most prominent approaches within this category is \emph{contrastive learning}, which uses the \textit{InfoNCE} loss \citep{oord2018representation} to align augmented views of the same image (positive samples) and differentiate between different images (negative samples). Well-known examples of contrastive learning frameworks include \textit{SimCLR}, \textit{MoCo}, and \textit{SwAV} \citep{chen2020simple, he_momentum_2020, caron_unsupervised_2021}.
    On the other hand, \emph{distillation-based methods} like \textit{BYOL}, \textit{DINO}, and \textit{DINOv2} \citep{grill2020bootstrap, caron_emerging_2021, oquab_dinov2_2024} eliminate the need for negative samples by training a student network using "soft labels" from a teacher network.

\paragraph{Generative Self-Supervised Learning}

    While computer vision has traditionally focused on discriminative learning approaches, the introduction of ViTs in 2021 marked an important shift in self-supervised representation learning \citep{balestriero_cookbook_2023}. By representing images as patch tokens,analogous to word tokens in natural language processing, ViTs enabled the transfer of language-based pretraining strategies to vision. In natural language processing, transformers are commonly trained with Masked Language Modeling (MLM), where parts of the input text are hidden and must be predicted. This inspired Masked Image Modeling (MIM), where subsets of image patches are masked or altered and the model learns to reconstruct them rather than solely distinguishing between samples. MIM has since driven substantial advances in generative self-supervised methods for vision.
    
    One of the earlier notable MIM approaches is \textit{BEiT}, which applies vector quantization to map continuous image patches to discrete visual tokens. Unlike approaches that predict pixel values directly, BEiT predicts these discrete tokens \citep{bao_beit_2022}. BEiT initially relied on an externally pre-trained tokenizer, but \textit{iBOT} improved the method by introducing an online tokenizer, which improved efficiency\citep{zhou_ibot_2022}.

    Another recent but popular approach within generative SSL is \textit{Masked Autoencoder} (\textit{MAE)}, which employs an asymmetric autoencoder framework. MAE’s unique feature is its encoder, which processes only unmasked patches of an image while ignoring masked ones.  A lightweight decoder is then tasked with reconstructing the original image from these encoded representations and the masked patches. By focusing only on reconstructing the missing (masked) parts, the decoder encourages the model to learn meaningful visual representations. This design not only reduces computational overhead but also ensures the model captures semantic content efficiently, making MAE faster and more effective than previous generative frameworks \citep{he2021masked}.

    More recently, the image-based implementation of the Joint Embedding Predictive Architecture (\textit{I-JEPA}) \citep{assran2023self} offers a novel SSL approach by predicting embeddings of masked regions based on context from unmasked regions within the same image. While similar in concept to MAE, I-JEPA predicts representations in the embedding space rather than the image space. Operating in the embedding space reduces sensitivity to pixel-level details typical of generative approaches and improves both semantic understanding and computational efficiency.

    \paragraph{Multimodal Self-Supervised Learning}

    While most self-supervised learning objectives are designed for single-stream data, multimodal FMs that handle both vision and language require specialized approaches to manage these diverse data types. Combining multiple modalities can be handled at various stages in the FM pipeline, a spectrum known as early (or feature level) to late fusion (or decision level).
    
    \textit{Late fusion} involves using separate encoders for each data type, each with its own SSL objectives. These modalities are processed independently and combined later in the pipeline, as seen, for example, in \cite{alayrac2022flamingo}. This method offers flexibility, allowing for the tailored processing of each modality and simplified integration at a later stage.
    
    In contrast, \textit{early fusion} integrates features from different modalities before or during the representation learning process. This results in a unified representation space that captures interactions between data types. Early fusion models, such as CLIP \citep{radford2021learning} and ALIGN \citep{jia2021scaling}, often utilize Image-Text Contrastive Learning (ITC). ITC aligns paired images and texts by maximizing their similarity and minimizing the similarity between mismatched pairs, effectively representing related multimodal data closely in a shared embedding space. Additionally, ITC can leverage unpaired data by aligning images with separate texts describing similar concepts, using techniques such as self-distillation \citep{bannur2023learning}.

    Finally, \textit{mid-fusion}, and \textit{multi-layer fusion} approaches, fall in between early and late fusion. These methods combine modalities at different preprocessing stages, often by creating layers within the neural network specifically designed for data fusion purposes \citep{nagrani2021attention}.
    
    The choice between early and late fusion methods depends on the specific requirements of the task: early fusion captures fine-grained interactions between modalities, but may involve higher dimensionality. Late fusion allows for greater flexibility in unimodal encoder pipeline design, but may fail to capture complex interaction patterns. Mid fusion offers a balance by integrating modalities at specific stages, though it can lead to increased model complexity.
    
\subsubsection{Adapting to Downstream Tasks}

Foundation models can generalize across a wide range of downstream tasks, such as image classification, object detection, segmentation, and question answering. Ideally, the model’s latent space is well-structured after pretraining, with the learned embeddings capturing relevant features for various tasks. Methods like PCA and k-nearest neighbors (k-NN) can be used to evaluate how well the embeddings cluster task-specific data, serving as indicators of how well the pretrained model may perform on certain tasks without further adaptation. If the embeddings are well-structured, simple methods like k-NN can yield satisfactory results. However, in most cases, additional adaptation is required to fine-tune the model for specific tasks. Adaptation methods can be viewed along a spectrum of increasing resource requirements: from approaches that leave model parameters untouched (prompting, linear probing), to those that add lightweight task-specific components while keeping the foundation model frozen (task-specific heads, parameter-efficient fine-tuning), and finally to approaches that update all parameters (full fine-tuning). The choice of adaptation method depends on factors like annotated data availability, computational resources, and the performance requirements for the task.

\paragraph{Prompt Engineering}
Prompt engineering is a lightweight adaptation technique that modifies a model’s behavior by carefully designing its inputs rather than altering its parameters. It is only applicable to language and vision-language models because it relies on the model’s ability to interpret natural language or structured textual inputs. Unlike models that require parameter updates for adaptation, prompt-based methods guide the model using carefully crafted input prompts, such as questions, commands, or structured text-based markers. For instance, a prompt might specify a task using natural language (e.g., "Find tumors in this MRI") or special tokens that indicate task-specific operations (e.g., "[SEGMENT] Brain tumor" in a vision-language model).

Prompting can be categorized into \textit{zero-shot} and \textit{few-shot} approaches. In zero-shot prompting, the model is given only task instructions without examples. In contrast, \textit{few-shot prompting} provides a handful of demonstration examples to illustrate the desired behavior before requesting a response. A key technique within few-shot prompting is \textit{In-Context Learning (ICL)}, introduced in GPT-3 \citep{brown_language_2020}. ICL involves providing examples, known as \textit{demonstrations}, within the input prompt to help the model understand and perform the task at hand. The examples are formatted as input-output pairs, which the model uses to infer the desired behavior and generate appropriate responses for new inputs \cite{dong2022survey, min2022rethinking}. Unlike traditional training with explicit labels, ICL helps the model learn from context without retraining, enabling it to generalize to new, unseen tasks. While most ICL methods are in few-shot setting, recently, many-shot ICL was also proposed and shown to improve performance \citep{agarwal2024many}. 

While classic prompt engineering applies mainly to language and vision-language models, there has been emerging research into visual prompting \citep{jia2022visual, bahng2022exploring}. However, unlike textual prompts, visual prompting techniques often require modifying some model parameters. These approaches typically introduce learnable visual tokens, small trainable modules, or perturbations to the input images that require additional optimization. As such, visual prompting is closer to parameter-efficient fine-tuning than to pure textual prompting, where the model remains entirely unchanged.

\paragraph{Linear Probing}
 Linear probing is a lightweight adaptation method where a simple linear classifier or regression model is trained on top of the embeddings generated by a pretrained foundation model. This involves either adding a linear layer for classification tasks, or applying regression on the embeddings for continuous value prediction. The pretrained model’s weights are frozen, and only the final layer is fine-tuned. Linear probing is computationally efficient and leverages the pretrained model’s embeddings, which, if well-structured, can often perform well with minimal updates. However, if the downstream task requires more complex decision boundaries or domain-specific features not captured by the pretrained model, additional adaptation methods may be needed.

\paragraph{Task-Specific Heads}
If linear probing is insufficient for capturing the complexity of the downstream task, a more powerful adaptation approach involves adding task-specific heads on top of the pretrained model. These smaller neural network modules are fine-tuned for specific tasks, while the foundation model’s weights remain frozen. Task-specific heads can be categorized into those for discriminative tasks, such as image classification and object detection, and generative heads, for tasks like image synthesis or text-to-image generation. For discriminative tasks, these heads may range from simple neural networks like MLPs and CNNs to more complex structures, including additional transformers. Generative heads, on the other hand, may involve autoregressive models, variational autoencoders (VAEs), or diffusion models \citep{kingma2022autoencoding, ho2020denoising}. These methods require supervised training and labeled data, but the amount of annotated data needed is significantly lower than what would be required to train an entire model from scratch. By focusing fine-tuning efforts on these smaller, task-specific components, computational costs can be reduced.

However, while task-specific heads are useful for many tasks, they may not always be sufficient. This is because these heads do not interact with or alter the core structure of the pretrained model; they only add task-specific modules on top of the model’s frozen layers. For some complex tasks, especially those requiring deep, domain-specific adjustments, task-specific heads might not provide the flexibility needed. In such cases, more advanced methods that allow modification of the model’s internal representations or layers are necessary.

\paragraph{Parameter-Efficient Fine-Tuning (PEFT)}

For scenarios where the pre-trained model needs to be finetuned, but full fine-tuning is too costly, Parameter-Efficient Finetuning (PEFT) techniques aim to reduce computational resources by fine-tuning only a specific set of model parameters \citep{tay2023efficient}. One of the most popular techniques is LoRA (Low-Rank Adaptation), which focuses on updating a set of low-rank matrices \citep{hu2021lora, zhu2023melo, lian2024less}. Another widely used PEFT method is adapter tuning, where small adapter modules are inserted into the layers of a pretrained model and trained for specific tasks while keeping the original model parameters frozen \citep{houlsby2019parameter}. This allows for efficient task-specific adaptation without requiring full model updates. Adapter tuning is particularly useful in multitask learning, where a single model is adapted to multiple tasks without retraining the entire network. Another approach is prompt tuning, where a small set of trainable prompt embeddings is optimized while keeping the model’s weights frozen \citep{lester2021power}. Unlike traditional prompt engineering, which involves manually crafting text-based prompts, prompt tuning learns task-specific soft prompts in a data-driven manner, making it a more scalable adaptation method.

\paragraph{Full Fine-Tuning} 

Full fine-tuning involves updating all of the model’s parameters based on the labeled data for a specific downstream task, allowing the model to adjust both its internal representations and the final task-specific layers. This approach typically leads to the best performance for the task at hand but comes with higher computational costs. Additionally, sufficient labeled data is required to prevent overfitting and ensure that the model generalizes well to new examples \citep{howard2018universal}.

Finally, some specialized techniques that involve full fine-tuning have been developed to improve task-following and alignment with human preferences. For instance, \emph{instruction-tuning} uses demonstration-based instructions during the model’s training phase, updating its parameters to optimize it for better task-following at inference time \citep{wei2021finetuned}. A vision-language extension of this approach is \textit{visual instruction tuning}, as implemented in LLaVA \citep{liu2023visual}, where vision-language models are fine-tuned using multimodal instruction-response pairs to improve their ability to follow natural language prompts grounded in images.
Furthermore, \emph{Reinforcement Learning from Human Feedback (RLHF)} \citep{kaufmann2023survey} incorporates human evaluations by training a reward model on preference data, which then guides reinforcement learning to align model outputs with human preferences. This enables more nuanced, context-dependent behaviors that would have been difficult to capture with standard supervised objectives.
Both instruction tuning and RLHF are also central to developing more \textit{agentic} capabilities in models: meaning, the capacity for autonomous planning and the execution of multi-step tasks \citep{acharya2025agentic}. These techniques provide the necessary alignment with demonstrations and human preferences to support such goal-directed behavior. When combined with \textit{chain-of-thought prompting} \citep{wei2022chain}, where the model explicitly generates intermediate reasoning steps, agentic FMs can perform complex multi-step reasoning, plan sequences of actions, and maintain context across tasks.

\section{Pathology}\label{section:pathology}

Pathology involves distinguishing diseases based on clinical, macroscopic, microscopic, and molecular characteristics by comparing them with predefined criteria. A pathology report provides a diagnosis but also includes information on prognostic (disease behavior) and predictive factors (response to therapy) \citep{funkhouser2020pathology}. Computational pathology has incorporated many downstream tasks from this workflow and continues to develop improved solutions for them using different methods.
Central to these tasks are whole slide images (WSIs), which are widely used to train AI models to improve cancer diagnosis accuracy, discover new biomarkers, and tailor treatment strategies according to individual patient profiles and specific tumor characteristics \citep{song2023artificial}. WSIs are obtained by digitizing conventional pathology glass slides using slide scanners, which contain thin tissue sections typically stained with hematoxylin and eosin (H\&E) or immunohistochemistry (IHC) stains. During this scanning process, the entire slide is captured at different magnification levels, providing a comprehensive view of the tissue sample, from a broad overview to the fine details of individual cells. The gigapixel size of the resulting WSIs makes it difficult for current hardware to process them entirely due to computation and memory constraints. Therefore, WSIs, at a fixed Microns Per Pixel (MPP), are usually split into smaller tiled patches, which are fed to the model individually.

Several factors can influence image analysis on WSIs. One major factor is the manual processing during preanalytical steps in a pathology laboratory, as it can introduce artifacts due to slight procedural differences \citep{taqi2018review}. Additionally, variations in H\&E staining protocols across laboratories and using different slide scanners contribute to color inconsistencies between WSIs from different centers \citep{ochi2024registered}, highlighting the need for robust models to overcome these variances.

\subsection{Towards Pathology Foundation Models} \label{section:towards_path_fm}
\paragraph{A two-stage pipeline} In digital pathology image analysis, deep learning methods typically follow a two-stage pipeline: a pre-trained backbone encoder computes lower-dimensional embeddings for individual tiles, which are then processed by a task-specific head. Segmentation tasks are an exception: nnU-Net models are commonly applied at the patch level, and encoder weights are often kept frozen to prevent overfitting when patch-level labels are limited. In contrast, slide-level prediction tasks, such as predicting clinical outcomes or molecular profiles, typically involve large sets of patches with only one label per slide. These tasks are framed as weakly supervised problems, where \textit{multiple-instance learning (MIL)} is used to aggregate patch-level embeddings into a slide-level prediction. Freezing encoder weights in this context also helps reduce computational demands, since end-to-end training from patches to slide-level predictions can be memory-intensive.

\paragraph{From transfer learning to general-purpose in-domain pre-training} Initially, transfer learning was employed to initialize encoders with weights from natural image datasets like ImageNet \citep{saillard2020predicting}. However, recent efforts have shifted toward SSL in the pathology domain, unlocking the potential for in-domain pretraining and significantly improving performance. Some studies applied contrastive SSL frameworks \citep{lu2019semi, dehaene2020self, zhao2020predicting}, while others explored pathology-oriented pretext tasks based on WSI characteristics like magnification or cross-stain prediction \citep{koohbanani2021self, yang2021self}. Early pre-training often focused on tissue types similar to downstream tasks, but more diverse pan-cancer datasets are now being used to train general-purpose encoders. These encoders serve as building blocks for various pathology tasks, with both public datasets like TCGA \cite{weinstein2013cancer} and proprietary data being used in this trend of scaling up data and model size.

\paragraph{Defining Pathology Foundation Models}
In this review we will use this notion of general-purpose in-domain pre-training to draw a line between what we define as foundation model and what we consider to be early advances of SSL for medical imaging.

\subsection{Pathology Foundation Models}

This section reviews recent advances in FMs for pathology, focusing on tile-level foundation models, which follow the two-stage pipeline outlined in Section \ref{section:towards_path_fm}. We first discuss the development of tile-level models, grouped by SSL frameworks. Next, we cover vision-language FMs, slide-level FMs, and the application of SAM for pathology data. An overview of landmark papers is provided in Tables \ref{table:pathology_vision}, \ref{table:pathology_vision-language} and \ref{table:pathology_slidelevel}.

\subsubsection{Contrastive Learning} \label{path_contrastive_learning}

\paragraph{SimCLR} \cite{ciga2022self} were the first to utilize a large number and diverse set of histopathology images for self-supervised training. Public data from TCGA, CPTAC\footnote{ \url{https://cptac-data-portal.georgetown.edu/cptacPublic}}, and several public challenge datasets were merged into a unified pretraining set with 25 thousand full WSIs and more than 206 thousand image patches, including different organs, staining types, and resolutions. Downstream performance on several classification, regression, and segmentation tasks revealed that in-domain pretraining of CNNs with SimCLR on a larger and more diverse dataset could improve the quality of the learned embeddings. 

\cite{azizi_robust_2023} used large-scale supervised transfer learning on natural images to initialize the backbone encoder before finetuning it on domain-specific unlabeled data using SimCLR. This training strategy, dubbed \textit{REMEDIS}, achieved better performance on tasks such as metastasis detection and survival prediction and more data-efficient generalization to out-of-distribution data compared to supervised baselines pre-trained on natural images only.

\paragraph{Other approaches} A challenge in contrastive learning is defining informative contrastive pairs. Typically, two augmented views of the same instance are considered a positive pair, and those from different instances are negative pairs. In histopathology, where patches of WSIs are often semantically similar, this approach can be problematic. To address this, \cite{wang2023retccl} proposed cluster-guided contrastive learning (CCL) to reduce false-negative samples. \cite{wang2022transformer} introduced semantically relevant contrastive learning (SRCL), an SSL strategy based on MocoV3 that uses latent space similarity to identify additional positive pairs. Using SRCL, a hybrid backbone integrating a CNN and multi-scale Swin Transformer architecture, coined \textit{CTransPath}, was trained on a large set of 32 thousand public pan-cancer WSIs from TCGA and PAIP\footnote{\url{http://www.wisepaip.org/paip}}. CTransPath was one of the first transformer-based feature extractors optimized using SSL at such a scale.

\subsubsection{Self-distillation with DINO} 
Several studies later adopted the DINO self-distillation framework for ViT-based pathology FMs. \cite{chen2022scaling} introduced the Hierarchical Image Pyramid Transformer (HIPT), which models the hierarchical structure of WSIs by aggregating visual tokens across different scales using DINO. This enables the capture of coarser-grained features across a larger context. \cite{campanella2023computational} applied DINO to a proprietary dataset with 3 billion patches from 432,000 WSIs, substantially larger than those in previous studies. \cite{kang2023benchmarking} benchmarked four representative SSL methods, including DINO, and found that in-domain pre-training consistently outperformed supervised pre-training with ImageNet, though no SSL method showed a significant advantage. \cite{pohjonen2024histoencoderdigitalpathologyfoundation} used DINO to train \textit{HistoEncoder}, an FM for prostate cancer that utilizes the XCiT backbone \citep{elnouby2021xcitcrosscovarianceimagetransformers}, reducing complexity compared to traditional ViTs. \textit{HistoEncoder} outperformed models pre-trained on natural images for cancer classification and mortality prediction, even without fine-tuning or with less data, making it more accessible to users with limited resources.

\subsubsection{Masked Image Modeling}
Inspired by the success of MIM in the natural image domain, most recent studies on pathology FMs use this SSL paradigm for in-domain pre-training. While \cite{yang2024foundation} introduced a BEiT-based pathology FM, the iBOT and DINOv2 frameworks are most commonly employed by other works that we discuss in more detail below.

\paragraph{iBOT} \cite{filiot2023scaling} analyzed the scalability of iBOT and the impact of data diversity by pretraining different ViT model sizes on either colon adenocarcinoma data or (a selection of) pan-cancer data from TCGA, which eventually resulted in the FM known as \textit{Phikon}. A large panel of 17 slide-level and tile-level downstream tasks, including the prediction of histological and molecular subtypes, genomic alterations, and survival, was used for evaluation. Phikon improved performance over previously published methods on all patch-level tasks and most slide-level classification tasks, with average gains of +1.4\% over CTransPath and +4.0\% over HIPT. These improvements were partly attributed to iBOT’s masked image modeling SSL method, which enabled richer representations than the contrastive (CTransPath) and distillation-based (HIPT) approaches.


\begin{table*}[t]
\caption{Tile-Level Vision Encoder Pathology FMs. Backbone and SSL objective show how the model was pretrained. The data size for pre-training is given in number of patches and the number of WSIs in parentheses. Availability shows whether the weights and pretraining data are available online.}
\label{table:pathology_vision}
\footnotesize
\begin{center}
\begin{tabular}{lllllll}
\toprule
\multirow{2}{*}{\textbf{Model}} & \multirow{2}{*}{Backbone} & \multirow{2}{*}{SSL} & \multicolumn{2}{c}{Pre-Training Data} & \multicolumn{2}{c}{Availability} \\
\cmidrule(lr){4-5}\cmidrule(lr){6-7}
& & & Size & Source & Weights & Data \\ 
\midrule
\textbf{REMEDIS} & ResNet-152 & SimCLR & 50M (29K) & TCGA & \cmark & \cmark \\
\textbf{RetCCL} & ResNet-50 & CCL & 15M (32K) & TCGA, PAIP & \cmark & \cmark \\
\textbf{CTransPath} & SwinTransformer & SRCL & 15M (32K) & TCGA, PAIP & \cmark & \cmark \\
HIPT & ViT-HIPT & DINO & 104M (11K) & TCGA & \cmark & \cmark \\
\textbf{Campanella et al.} & ViT-S & DINO & 3B (400K) & MSHS\textsuperscript{a} & \xmark & \xmark \\
\textbf{Lunit} & ViT-S/8, ViT-S/16 & DINO & 33M (37K) & TCGA, TULIP\textsuperscript{a} & \cmark & \cmark/\xmark\textsuperscript{c} \\
\textbf{Phikon} & ViT-B/16 & iBOT & 43M (6K) & TCGA & \cmark & \cmark \\
\textbf{UNI} & ViT-L/16 & DINOv2 & 100M (100K) & Mass-100K\textsuperscript{a} & \cmark & \xmark\textsuperscript{b} \\
\textbf{Virchow} & ViT-H/14 & DINOv2 & 2B (1.5M) & MSKCC\textsuperscript{a} & \cmark & \xmark \\
\textbf{RudolfV} & ViT-L/14 & DINOv2 & 1.2B (134K) & TCGA, Proprietary & \xmark & \xmark \\
\textbf{PLUTO} & ViT-S/8, ViT-S/16 & Modified DINOv2 & 195M (158K) & TCGA, PathAI\textsuperscript{a} & \xmark & \xmark \\
\textbf{Hibou-L} & ViT-L/14 & DINOv2 & 1.2B (1.1M) & Proprietary & \cmark & \xmark \\
\textbf{H-optimus-O} & ViT-G & DINO2 & - (500K) & Proprietary & \xmark & \xmark \\
\textbf{Virchow2G} & ViT-G/14 & DINOv2 & 1.9B (3.1M) & MSKCC\textsuperscript{a} & \xmark & \xmark \\
\textbf{Phikon-v2} & ViT-L & DINOv2 & 456M (58.4K) & PANCAN-XL & \cmark & \cmark/\xmark\textsuperscript{c} \\
\textbf{Atlas} & ViT-H/14 & RudolfV (DINOv2-based) & 520M (1.2M) & Proprietary & \xmark & \xmark \\
\bottomrule
\end{tabular}
\end{center}

\begin{minipage}{\textwidth}
\footnotesize
\textsuperscript{a} Proprietary data \quad
\textsuperscript{b} Available upon request \quad
\textsuperscript{c} Partly available
\end{minipage}
\end{table*}

\paragraph{DINOv2} 
Around the same time, \cite{chen2024towards} introduced \textit{UNI}, a ViT-Large model trained with DINOv2 on the Mass-100k dataset, which included over 100,000 WSIs across 20 pathology types. UNI was compared with CTransPath and REMEDIS on a range of tasks, including slide-level classification and ROI-level tasks (classification, segmentation, retrieval, and prototyping) across 34 anatomic pathology tasks. UNI outperformed other models on average but performed similarly to REMEDIS in ROI segmentation. UNI also demonstrated better label efficiency and resolution-agnosticity in classification tasks.

Shortly after, \citep{vorontsov2024foundation} published \textit{Virchow}, a ViT-Huge model trained on a proprietary dataset of 2 billion patches from 1.5 million WSIs and 119,000 patients, making it the largest pathology FM at the time. Virchow outperformed UNI, Phikon, and CTransPath on public benchmark tile-level tasks and pan-cancer detection. Though it performed better in biomarker prediction, the differences were not significant. Notably, in detecting prostate cancer, invasive breast cancer, and breast cancer metastasis to lymph nodes, Virchow was competitive with specialist commercial models, despite using a less refined training set.

Building on \citep{vorontsov2024foundation}, \cite{zimmermann2024virchow} introduced domain-inspired augmentations and regularization to improve Virchow’s training. The dataset was expanded to 3.1 million WSIs from 225,000 patients across various institutions and tissue types. The resulting \textit{Virchow2} model improved performance on in-distribution and out-of-distribution benchmarks, and further improvements were seen with the larger Virchow2G model, using a ViT-G architecture.
More recently, building on Phikon, \textit{Phikon-v2} was proposed, which was trained on 58.4K publically available WSIs with DinoV2 \citep{filiot2024phikon}. Phikon-v2 outperformed Phikon, and its performance was on-par with that of FMs trained on large proprietary datasets.

Besides the aforementioned studies, several other pathology FMs have been pre-trained with DINOv2, such as RudolfV \cite{dippel2024rudolfv}, PLUTO \citep{juyal2024pluto}, Hibou-B/Hibou-L \citep{nechaev2024hibou}, and H-optimus-O \citep{hoptimus0}.
Finally, \textit{Atlas} \citep{alber2025novel} is a ViT-H/14 pathology foundation model trained using a DINOv2-style self-supervised learning framework adapted from RudolfV. Unlike prior models that typically rely on H\&E-stained slides at a single magnification, Atlas was trained on a highly diverse dataset of 1.2 million WSIs from 490,000 cases, containing 70 tissue types, multiple magnifications, scanner types, and staining protocols, including H\&E, IHC, and special stains. Despite having a moderate size of 632 million parameters, which is the same as Virchow2 and significantly fewer than Prov-GigaPath’s 1.1 billion, Atlas achieves the highest average performance across 21 public benchmarks as well as additional tumor-microenvironment and cancer typing tasks. It outperforms both Virchow2 and H-Optimus-0 by 1.1\%, suggesting that broad data diversity combined with strong self-supervised learning might be favored over simply scaling in both model and dataset size (see also \ref{path_scale}).

\subsubsection{Vision-Language Foundation Models}

In addition to vision-only pathology FMs, research on multimodal learning from vision and language data in pathology has grown rapidly in recent years. As a result, several VLFMs, mostly pre-trained with cross-modal contrastive objectives, have been published \citep{huang2023visual, lu2023visual, ikezogwo2024quilt, lu2024visual, shaikovski2024prism,  sun2024pathasst, xu2024whole, zhang_biomedclip_2024, qu2024rise, zhang2023text}. The majority of these works use tile-level vision encoders trained with image-text pairs sourced from, for example, pathology textbooks, research papers, YouTube videos, and medical Twitter.

Models such as \textit{MI-Zero} and \textit{PLIP} extend the CLIP framework to pathology. \textit{MI-Zero} aligns patch-level embeddings from a vision encoder with text embeddings from a GPT-style transformer trained on pathology reports and PubMed abstracts, allowing zero-shot cancer subtyping without additional labels \citep{lu2023visual}. Similarly, \textit{PLIP}, trained on data curated from Twitter, performs well on tasks like cancer classification and disease progression prediction \citep{huang2023visual}. Building on this, \textit{QUILTNET} and \textit{CONCH} further increase the scale at which VLFMs are trained, with \textit{QUILTNET} fine-tuned on 1M image-text pairs (QUILT dataset) and excelling in cross-modal retrieval \citep{ikezogwo2024quilt}, and \textit{CONCH}, trained on 1.17M pairs using the CoCa framework, achieving strong results in classification, retrieval, and tissue segmentation tasks \citep{lu2024visual}.

\begin{table*}[t]
\caption{Vision-Language Pathology Foundation Models. Data size shows amount of image-text pairs. Since these are collected from non-clinical sources such as PubMed, Twitter or Youtube, no distinction between WSI and patches is made.}
\label{table:pathology_vision-language}
\footnotesize
\begin{center}
\begin{tabular}{@{}lllllll@{}}
\toprule
\multirow{2}{*}{\textbf{Model}} & \multirow{2}{*}{Backbone} & \multirow{2}{*}{SSL} & \multicolumn{2}{c}{Pre-Training Data} & \multicolumn{2}{c}{Availability} \\
\cmidrule(lr){4-7} 
& & & Size & Source & Weights & Data \\
\midrule
\textbf{PLIP}         & ViT-B/32 + Transformer & ITC                    & 208K      & OpenPath                     & \cmark & \cmark \\
\textbf{MI-Zero}      & CTransPath + GPT2-medium & ITC                  & - (33K)   & Proprietary                  & \cmark & \xmark \\
\textbf{QuiltNet}     & ViT-B/32 + PubmedBert & ITC                     & 34K       & QUILT                         & \cmark & \cmark \\
\textbf{CONCH}        & ViT-B/16 + GPT2-medium & iBOT\textsuperscript{a} + CoCa & 1.1M  & PMC-Path, EDU                 & \cmark & \cmark \\
\textbf{PathCLIP}     & CLIP vision encoder + Vicuna-13b & ITC          & 207K      & PathCap                       & \cmark & \cmark \\
\textbf{PA-LLaVA}     & PLIP vision encoder + Lama3 & ITC + ITM          & 1.4M      & PMV, PMC-OA, Quilt-1M        & \cmark & \cmark \\
\textbf{PathChat}     & Uni + LLama2           & CoCa                   & 100K      & PMC-OA, Proprietary           & \cmark/\xmark\textsuperscript{b} & \cmark/\xmark\textsuperscript{b} \\
\textbf{Quilt-LLaVA}  & QuiltNet vision encoder + Vicuna & ITC            & 723K      & Quilt-1M                       & \cmark & \cmark \\
\bottomrule
\end{tabular} 
\end{center}
\begin{minipage}{\textwidth}
\footnotesize
\textsuperscript{a} iBOT is used for training their own image encoder.  \quad
\textsuperscript{b} Dataset is partly available. Weights are only available trained on model with open-source data, but not for model trained on full dataset. \quad
\end{minipage}
\end{table*}

More recently, generative approaches to VLFMs for pathology have also emerged. For instance, \textit{PathCLIP}, a CLIP variant fine-tuned on 207K pathology image-caption pairs from PubMed, serves as the visual backbone for \textit{PathAsst}, a multimodal generative model combining PathCLIP with the Vicuna-13b language model \citep{sun2024pathasst}. \textit{PathAsst} enables advanced tasks such as visual question answering, multimodal dialogue, and image interpretation. Other instruction-tuned generative models, including \textit{PA-LLaVA} \citep{dai2024pa}, \textit{PathChat} \citep{lu2024multimodal}, and \textit{Quilt-LLaVA} \citep{seyfioglu2024quilt}, show similar strong performance in visual question  answering tasks.

\subsubsection{Slide-level Foundation Models}{\label{slide-level}}

The tile-level pathology FMs discussed thus far are scalable and computationally efficient, but their reliance on local tiles means they lose global spatial context, limiting performance on tasks that require holistic assessment of tissue architecture. Some hierarchical tile-level models, such as \textit{HIPT}, partially address this limitation by aggregating multi-scale information, extending the effective context window. However, \textit{HIPT }still operates primarily at the patch level and cannot capture truly global slide-level patterns in a single self-supervised model.

Slide-level approaches address this by directly modeling entire WSIs, aiming to capture both local and global dependencies. One such method is the vision encoder \textit{LongVit }\citep{wang2023image}, which leverages DINO and LongNet \citep{ding2023longnet} to process the extremely long token sequences produced by gigapixel images, enabling end-to-end slide-level representation learning. While LongVit can model a larger context than conventional tile-level encoders, slide-level self-supervised modeling remains computationally demanding and less widely benchmarked than tile-level approaches.

Recent models have combined patch- and slide-level encoders with vision-language alignment to enhance feature generalization and enable multimodal tasks. For example, \textit{Prov-GigaPath }\citep{xu2024whole} uses a vision transformer (DINOv2) for tile-level features and a LongNet model to capture slide-level features, with CLIP-based vision-language alignment improving zero-shot performance in cancer subtyping and gene mutation prediction. This model outperforms patch-level FMs like REMEDIS and UNI in cancer detection and tumor origin prediction. Similarly, \textit{PRISM} \citep{shaikovski2024prism} aggregates tile-level embeddings from Virchow to generate slide-level embeddings for zero-shot cancer subtyping, detection, and clinical report generation.
\textit{CHIEF} \citep{wang2024pathology} builds on this framework by pretraining a patch-level encoder, CTransPath, and combining it with weakly supervised pretraining on 60K WSIs. It also uses CLIP-based text prompts to encode anatomical site information, improving generalization for cancer detection and tumor origin prediction. \textit{TITAN} \citep{ding2024multimodal} integrates histology knowledge distillation with contrastive vision-language learning to align regions of interest with captions and pathology reports, excelling in cancer subtyping, rare disease classification, and multimodal tasks like report generation.

Other models enhance slide-level representation learning by integrating additional biological context or modalities.\textit{ KEEP }\citep{zhou2024knowledgeenhancedpathologyvisionlanguagefoundation} incorporates a disease knowledge graph to enhance performance in rare cancer detection and subtyping. \textit{mSTAR }\citep{xu2024multimodal} integrates WSIs, pathology reports, and RNA-Seq data, significantly improving unimodal and multimodal tasks, particularly genomic information use. \textit{THREADS} \citep{vaidya2025molecular} introduces molecular supervision by aligning whole-slide image features with matched transcriptomic and genomic profiles, enabling representations that reflect both morphological and molecular characteristics. Despite having a smaller slide encoder than those of PRISM and GigaPath, THREADS achieves superior performance on oncology tasks, including rare event prediction, and outperforms lightweight models like CHIEF.

Finally, \textit{COBRA} \citep{lenz2024unsupervised} represents an ensemble method that aggregates patch embeddings from multiple foundation models in feature space using a contrastive pre-training strategy. Tested on FMs CTransPath, UNI, Virchow2, and H-Optimus-0, COBRA is flexible and applicable to any ensemble, demonstrating strong performance despite being pre-trained on only 3048 WSIs.

\begin{table*}[t]
\caption{WSI-Level Pathology Foundation Models. Type refers to whether the FM is a vision model (VM) or a vision-language model (VLM). Backbone and SSL objective show how the model was pretrained. The pre-training dataset size is given in number of patches and the number of WSIs in parentheses. Availability shows whether the weights and pretraining data are available online.}
\label{table:pathology_slidelevel}
\footnotesize
\begin{minipage}{\textwidth} \begin{center}
\centerline{\begin{tabular}{@{}llllllll@{}}
\toprule
\multirow{2}{*}{\textbf{Model}} & \multirow{2}{*}{Type} & \multirow{2}{*}{Backbone} & \multirow{2}{*}{SSL} & \multicolumn{2}{c}{Pre-training Data} & \multicolumn{2}{c}{Availability} \\
\cmidrule(lr){5-8} 
& & & & Size & Source & Weights & Data \\
\midrule
\textbf{Giga-SSL}       & VM  & ResNet-18                 & SimCLR                     & - (12K)     & TCGA                & \cmark & \cmark \\
\textbf{LongViT}        & VM  & LongNet                   & DINO                       & 1M (10K)    & TCGA                & \cmark & \cmark \\
\textbf{PRISM}          & VLM & Virchow + BioGPT          & CoCa                       & - (590K)    & Proprietary         & \cmark & \xmark \\
\textbf{Prov-GigaPath}  & VLM\textsuperscript{a} & ViT-G + PubMedBERT         & DINOv2 + ITC               & 1.3B (170K) & Prov-Path           & \cmark & \xmark \\
\textbf{mSTAR}          & VLM & ViT-L                     & mSTAR                      & - (26K)     & TCGA                & \xmark & \xmark \\
\textbf{KEEP}           & VLM & Uni + PubMedBERT          & Knowledge-enhanced ITC     & - (143K)    & OpenPath, Quilt-1M  & \cmark & \cmark \\
\textbf{CHIEF}          & VLM\textsuperscript{a} & CTransPath + CLIP text encoder & SRCL (v)+ ITC (vl) & 15M (61K)   & Proprietary         & \cmark & \xmark \\
\textbf{TITAN}          & VLM\textsuperscript{a} & ViT-B + CONCHv1.5          & iBOT (v) + CoCa (vl)       & - (60K)     & GTEx                & \cmark & \cmark \\
\textbf{COBRA}          & VLM & ABMIL + Mamba-2           & COBRA                       & - (3K)      & mTCGA               & \cmark & \cmark \\
\textbf{THREADS}        & VM  & CONCHV1.5 (patch) + cGPT (gene) & Contrastive           & - (47K)     & MBTG-47K            & \xmark\textsuperscript{b} & \cmark/\xmark\textsuperscript{c} \\
\bottomrule
\end{tabular}} \end{center}

\footnotesize 
\textsuperscript{a} Authors train both an image encoder as well as a VL-alignment module. 
\textsuperscript{b} Planned to be released. 
\textsuperscript{c} Partially available: TCGA and GTEx are open-source, but pretraining data from BWH and MGH are proprietary.
\end{minipage}
\end{table*}

\subsubsection{Applying SAM in Pathology}

The \textit{Segment Anything Model} (SAM) is a vision foundation model originally designed for universal image segmentation, trained on over a billion masks from natural images \citep{kirillov2023segment}. Its flexibility in generating segmentation masks from simple prompts makes it appealing for medical applications, even though it does not fully align with the self-supervised pretraining paradigm central to our review.

Consequently, SAM has been explored either via direct zero-shot use on histopathology images \citep{chauveau2023segment, deng2023segment, zhang2023input, zhu2023samms} or through in-domain fine-tuning on pathology-specific datasets \citep{ranem2023exploring, zhang2023sam, chen2024sam, cui2024all, su2024adapting, zhang2024sam2}. These works show that SAM can deliver promptable segmentation without extensive annotation. However, its performance remains limited in addressing histopathological complexities, varying scales, and dense cellular structures, making SAM most useful as a transfer baseline rather than a domain-specific foundation model.

\subsection{Concluding Remarks}

Research on SSL and FM development in pathology has made great progress in recent years and has proven fruitful in several digital pathology applications. Early models like \textit{CTransPath} laid the groundwork for domain-specific adaptations, while later efforts such as \textit{Virchow} introduced scalability and multi-task capabilities at the slide level. The focus has shifted toward scaling both data and model size, as seen, for example, in the performance of \textit{UNI} and \textit{Phikon}, which adopt broader VFM frameworks with specific adaptations for cancer and pathology tasks. 
Vision foundation models have been at the forefront of these efforts, with notable success in improving performance across various pathology tasks. However, there have also been works using vision and language, such as using vision-language alignment to enhance whole-slide FMs and developing visual question answering FMs for pathology.

Despite these advancements, challenges remain. Scaling datasets and model sizes has brought notable performance improvements. Still, results suggest a plateau in gains, emphasizing the need for better data curation, particularly to represent rare pathologies and diverse staining protocols. Additionally, foundational algorithms optimized for natural images may require further adaptation to fully exploit the unique characteristics of histopathology data, including high resolution and spatial redundancy. 
Tile-level FMs have so far proven more scalable and robust, as they can be trained efficiently on individual patches, require lower computational resources than slide-level models, and have demonstrated strong performance across diverse datasets and benchmark tasks.
 In contrast, slide-level FMs promise richer contextual understanding by modeling entire WSIs, but remain limited by their computational demands, smaller training cohorts, and fewer established benchmarks.
Future work will require balancing scale with task specificity while addressing technical hurdles such as standardizing data across institutions.

\section{Radiology}\label{section:radiology}

Radiology data is obtained through various imaging techniques, such as plain radiography (X-ray) and computed tomography (CT), magnetic resonance imaging (MRI), ultrasound (US), and nuclear medicine imaging (such as PET scans) \cite{islam2023introduction}. Each modality provides data in different formats and dimensions: X-rays and CT scans use the same ionizing radiation to produce two-dimensional images and three-dimensional volumetric images, respectively, enabling for example detection of fractures, infections, tumors, and blood clots. MRI uses magnetic fields and radiofrequency waves to generate detailed 3D images of soft tissues, ideal for examining the brain, spine, muscles, and joints. Ultrasound uses high-frequency sound waves for 2D and 3D imaging, commonly used in prenatal check-ups and cardiac assessments. Nuclear medicine employs radioactive tracers to evaluate organ function, which is particularly useful for diagnosing cancer and cardiovascular and brain diseases. X-rays, CT, MRI, and US scans are the most frequently performed imaging tests in clinical practice.

The final output of a radiology exam in clinical practice is usually a written radiology report. This report describes the radiologist's interpretation of a particular scan, and typically summarizes key observations such as the size, shape, and location of abnormalities, as well as diagnoses, the severity of the condition, and any recommendations for additional tests or treatments. If prior scans are available, observations are related to earlier time-points and changes are reported.
These reports can therefore provide a rich source of information for AI models, as they can essentially be used as a guide for identifying areas of interest within the scan.
However, radiology reports are usually not standardized. Variability can arise from differences in terminology, reporting formats, and levels of detail among radiologists and institutions. Some reports may follow structured templates, while others might be more free-form, leading to inconsistencies in how information is conveyed. Leveraging the combination of the abundance of visual data with paired unstructured text is what makes radiology an excellent candidate for use of VLFMs. However, the textual reports are often created based on changes between two imaging scans, making CLIP-like strategies without severe report processing challenging.

\subsection{Early Advances in Radiology Models}
Unlike pathology, where the development of foundation models has seen significant progress, radiology has seen relatively fewer developments, with less foundation models emerging. Early radiology models primarily focused on leveraging transformer architectures for tasks like report generation and segmentation using CNN-transformer hybrids such as \cite{xie_cotr_2021}, \cite{hatamizadeh_unetr_2021}, and \cite{chen_transunet_2021}. However, these models did not utilize self-supervised learning, which limited their scalability. As research progressed, the focus shifted towards developing SSL techniques specialized for radiological imaging, particularly contrastive vision-language learning. 
Early work by \cite{feng_parts2whole_2020} improved feature learning by leveraging part-whole relationships. \cite{zhang2020contrastive} introduced \textit{ConVIRT} to align medical images with textual descriptions, followed by \textit{GLoRIA} \citep{huang_gloria_2021}, which refined this alignment at the sub-region level in radiology reports. Later, \cite{boecking2022making} focused on chest X-rays, enhancing semantic modeling through contrastive learning, which was further improved by \textit{Bio-ViL-T} \citep{bannur2023learning} with the addition of temporal dynamics, although the scale of these methods remained limited compared to modern foundation models.

\subsection{Foundation Models for Radiology}
This section reviews recent advances in FMs for radiology,
focusing on vision-language foundation models for a broad range of imaging modalities, VLFMs specifically developed for X-Ray, VLFMs developed for CT and MRI, which face challenges in extending to the 3D nature of these modalities, and vision FMs, which   - unlike in pathology -  are only recently becoming more studied. An overview of papers is also provided in Tables 4-6, which are split according to imaging modality: Table \ref{table:rad_x} lists the FMs developed for X-Ray, \ref{table:rad_3d} shows the FMs for CT and/or MRI and Table \ref{table:rad_gen} covers  multi-modality foundation models.

\subsubsection{Generalist Vision-Language Foundation Models}\label{rad:generalist}
As foundation models became more popular, within the medical domain, a category of models emerged that emphasizes scale by incorporating a wide variety of medical imaging modalities. These generalist vision-language models utilize extensive paired datasets to address various medical imaging modalities, including radiology, pathology and ophthalmology. Models such as \textit{PubMed-CLIP} \citep{eslami_does_2021}, \textit{PMC-CLIP} \citep{lin2023pmc}, \textit{BiomedCLIP} \citep{zhang_biomedclip_2024}, and \textit{MEDVInT} \citep{zhang_pmc-vqa_2023} utilized extensive datasets from PubMed to perform multimodal tasks like disease classification, report generation, image retrieval, and medical question answering (MedVQA). Models such as \textit{LLaVA-Med} \citep{li_llava-med_2023}, which finetuned LLaVA \citep{liu2023visual}, and \textit{Qilin-Med-VL} \citep{liu_qilin-med-vl_2023}, which used curriculum learning and instruction tuning, specialize in MedVQA tasks. Similarly, \textit{Med-Flamingo} \citep{moor_med-flamingo_2023}, built further on the trained  Flamingo architecture \citep{alayrac2022flamingo} for general medical VQA. 
Although evaluated on radiological tasks such as X-ray, these models were not specifically tailored for radiological imaging, focusing more broadly on multimodal medical datasets. Additionally, these models did not utilize 3D data, instead relying on 2D slices of 3D modalities such as CT and MRI.

\begin{table*}[t]
\caption{Foundation Models for Multi-Domain Radiology. Type refers to whether the FM is a vision model (VM) or a vision-language model (VLM). Backbone and SSL objective show how the model was pretrained. Imaging domain shows which medical image types were used to pre-train the model. Dataset sizes are given as the number of 2D images used for pre-training, unless specified otherwise. Availability shows whether the weights and pretraining data are available online.}
\label{table:rad_gen}
\footnotesize
\begin{minipage}{\textwidth} \begin{center}
\centerline{\begin{tabular}{@{}llllllll@{}}
\toprule
\multirow{2}{*}{\textbf{Model}} & \multirow{2}{*}{Type} & \multirow{2}{*}{Imaging Domain} & \multirow{2}{*}{Backbone} & \multirow{2}{*}{SSL Objective} & \multirow{2}{*}{Data Size} & \multicolumn{2}{c}{Availability} \\
\cmidrule(lr){7-8}
& & & & & & Weights & Data \\
\midrule
\textbf{PubMedCLIP}   & VLM & 2D Radiology\textsuperscript{c} & ResNet50 + ViT-B/32 & ITC & 80K & \cmark & \cmark \\
\textbf{UniMiSS}      & VM  & CT(3D), X-Ray         & MiT                  & MIM & 114K\textsuperscript{d} & \xmark & \cmark \\
\textbf{BiomedCLIP}   & VLM & Multimodal\textsuperscript{b}   & ViT-B/16 + PubMedBERT & ITC & 15M & \cmark & \cmark \\
\textbf{PMC-CLIP}     & VLM & 2D Radiology\textsuperscript{c} & ResNet50 + PubMedBERT & ITC + MLM & 1.65M & \cmark & \cmark \\
\textbf{MEDVInT}      & VLM & Multimodal\textsuperscript{b}   & ResNet50 + Transformer & MLM & 227K & \xmark & \cmark \\
\textbf{LLaVA-Med}    & VLM & Multimodal\textsuperscript{b}   & LLaVA                 & ITC & 600K & \cmark & \cmark \\
\textbf{LVM-Med}      & VM  & CT(2D), MRI, X-ray, US & ResNet-50 + ViT       & Graph Matching & 1.3M & \xmark & \cmark \\
\textbf{Med-Flamingo} & VLM & Multimodal\textsuperscript{b}   & ViT/L-14 + LLaMA-7B   & ITC & 1.6M & \cmark & \cmark \\
\textbf{RadFM}        & VLM & X-Ray, CT, MRI (3D), US, PET & 3D ViT + MedLLaMA-13B & Generative ITC & 16M\textsuperscript{e} & \cmark & \cmark \\
\textbf{Qilin-Med-VL} & VLM & Multimodal\textsuperscript{b}   & ViT/L-14 + LLaMA-13B  & ITC & 580K & \cmark & \cmark \\
\textbf{SAT}          & VLM & Multimodal\textsuperscript{b}   & 3D U-Net + BioBERT    & ITC & 302K & \xmark & \cmark \\
\textbf{VISION-MAE}   & VM  & CT, MRI, PET, X-Ray, US & Swin-T               & MAE & 2.5M & \xmark & \xmark \\
\textbf{RadCLIP}      & VLM & CT (2D+3D), X-Ray, MRI & ViT-L/14             & Contrastive & 1.2M & \xmark & \cmark \\
\bottomrule
\end{tabular}} \end{center}

\footnotesize  
(a) X-Ray, US, MRI, PET, fluoroscopy, mammography, angiography. 
(b) Multimodal = multiple radiology + pathology modalities. 
(c) Includes CT, MRI, PET, X-Ray, US, fMRI, etc. 
(d) Dataset includes 109K 2D images + 5K 3D volumes. 
(e) Includes 15.5M 2D and 500K 3D image-text pairs.
\end{minipage}
\end{table*}

\subsubsection{Chest X-Ray Vision-Language Foundation Models}
One subdomain within radiology that has seen relatively more FM related developments than other modalities is Chest X-ray (CXR). CXR lends itself well to foundation model research due to its widespread use in clinical settings, the availability of large and well-established datasets such as \cite{irvin2019chexpert, johnson2019mimic}, and its relatively reduced computational complexity compared to more intricate 3D modalities like CT and MRI. 
However, unlike the earlier mentioned generalist models that can scale by incorporating multiple imaging modalities, gathering large, paired CXR-specific datasets remains challenging. As a result, some models like \textit{MedCLIP} \citep{wang_medclip_2022} have addressed this by leveraging unpaired data with semantic matching, enabling the use of more diverse datasets. 
Other models focused on enhancing the quality of existing paired data to maximize the learning potential and improve feature extraction. For example, \textit{CheXzero} \cite{tiu_expert-level_2022} employed a contrastive learning approach using only the concise "impressions" section of radiology reports, leading to better feature extraction and zero-shot disease classification and \textit{CXR-CLIP} \citep{you_cxr-clip_2023} built on this by creating high-quality image-text pairs with radiologist-designed prompts, further improving feature retrieval and interpretability.
\textit{KAD} \cite{zhang_knowledge-enhanced_2023} incorporated a medical knowledge graph into its training process, allowing for deeper reasoning and improved visual representation learning, although its reliance on single-source data limited its adaptability. To overcome this, \textit{UniChest} \cite{dai_unichest_2024} introduced a conquer-and-divide framework that systematically captured shared patterns across diverse datasets while isolating source-specific variations, resulting in better generalization across CXR benchmarks. Similarly, \textit{MedKLIP} \cite{wu2023medklip} focused on fine-grained alignment between images and text to enhance interpretability and disease classification, while \textit{MaCo} \cite{huang_enhancing_2023} introduced a masked contrastive approach, refining image-text alignment and further improving zero-shot classification performance.

In addition to knowledge-enhancing models, others like \textit{ELIXR} \cite{xu2023elixr} and \textit{MAIRA-1} \cite{hyland2023maira} focused on efficient adaptation by using specialized adapters to align visual encoders with large language models, reducing the need for full retraining. A more recent adaptation of \textit{MAIRA-1}, \textit{MAIRA-2}, combines two pre-trained image- and language encoders and adapts them for report generation from chest X-ray images \citep{bannur2024maira}. More recently, Google’s \textit{CheXAgent} \cite{chen_chexagent_2024} marked a shift toward scaling with large datasets. By leveraging the CheXinstruct dataset, which includes 6 million instruction-image-answer triplets, CheXAgent focused on large-scale instruction-tuned learning and demonstrated superior performance across multiple clinically relevant tasks, highlighting the potential of large-scale, instruction-based approaches.

\begin{table*}[t]
\caption{Foundation Models specialized for Chest X-Ray. Type refers to whether the FM is a vision model (VM) or a vision-language model (VLM). Backbone and SSL objective show how the model was pretrained. Data sizes are given as the number of 2D images used for pre-training. Availability shows whether the weights and pretraining data are available online.}
\label{table:rad_x}
\footnotesize
\begin{minipage}{\textwidth} \begin{center}
\begin{tabular}{@{}lllllll@{}}
\toprule
\multirow{2}{*}{\textbf{Model}} & \multirow{2}{*}{Type} & \multirow{2}{*}{Backbone} & \multirow{2}{*}{SSL Objective} & \multirow{2}{*}{Data Size} & \multicolumn{2}{c}{Availability} \\
\cmidrule(lr){6-7}
& & & & & Weights & Data \\
\midrule
\textbf{BioViL-T} & VLM & Custom CNN–Transformer + BERT & Temporal Multi-Modal & 174.1K & \xmark & \cmark \\
\textbf{ELIXR} & VLM & SupCon + T5 & ITC & 220K & \xmark & \cmark \\
\textbf{MaCo} & VLM & ViT-B/16 + BERT & MLM & 377K & \xmark & \cmark \\
\textbf{CXR-CLIP} & VLM & ResNet50 + BioClinicalBERT & ITC & 15M & \cmark & \cmark \\
\textbf{UniChest} & VLM & ResNet50 + BioClinicalBERT & ITC & 686K & \cmark & \cmark \\
\textbf{RAD-DINO} & VM & ViT-B/14 & DinoV2 & 838K & \xmark & \cmark/\xmark \\
\textbf{CheXagent} & VLM & Ensemble & ITC + IC & 6.1M\textsuperscript{a} & \xmark\textsuperscript{b} & \xmark\textsuperscript{b} \\
\textbf{Ray-DINO} & VM & ViT-L & DinoV2 & 863K & \xmark & \cmark \\
\bottomrule
\end{tabular} \end{center}

\footnotesize  
(a) Does not provide pre-training dataset size, but finetunes on 6.1M samples. 
(b) Planned to be released.
\end{minipage}
\end{table*}

\subsubsection{Extending to 3D Modalities}
Unlike X-ray imaging, which is inherently 2D, CT and MRI are volumetric modalities that pose additional challenges for foundation models. Most FM architectures, such as vision transformers and contrastive vision–language frameworks, are optimized for 2D images. Applying them directly to 3D data requires either slicing volumes into 2D images, thereby losing important spatial context, or developing dedicated 3D encoders, which face obstacles including higher computational cost, limited availability of large-scale 3D datasets, and the difficulty of learning rich volumetric representations. As a result, early radiology FMs often defaulted to processing 3D scans as collections of 2D slices.

Early examples include \textit{CLIP-LUNG} \citep{lei2023clip}, which adapted the CLIP framework to CT but only at the slice level, thereby missing volumetric information. Subsequent works, such as \textit{M3FM} \citep{niu2023medical} and \textit{MedBLIP} \citep{chen2023medblip}, began to incorporate 3D data by dividing scans into 3D patches or sub-volumes, though these approaches still offered only partial use of spatial context.

More recent models have advanced toward fully volumetric approaches. \textit{RadFM} \citep{wu_towards_2023} was a notable milestone, introducing the MedMD dataset (16M 2D and 3D image–text pairs) and employing a 3D vision encoder. However, performance remained limited for vision-specific tasks like segmentation, partly due to the dominance of 2D data in the training set. Other efforts, such as \textit{M3D-LaMed} \citep{bai2024m3d} and \textit{SAT} \citep{zhao_one_2023}, focused specifically on 3D segmentation, while \textit{Rad-CLIP} \citep{lu2024radclip} and \textit{CT-CLIP} \citep{hamamci_foundation_2024} extended CLIP-style training to volumetric data. Among them, Rad-CLIP relied on pseudo-3D structures from 2D slices, whereas CT-CLIP used a 3D vision transformer and a dataset of 50K 3D chest CT volumes.
\textit{Merlin} \citep{blankemeier2024merlin} represents a notable step forward in handling 3D CT data, integrating structured EHR data and unstructured text. However, its focus on abdominal CT limits its applicability as a general-purpose foundation model.

Taken together, these works illustrate a gradual shift from 2D slice-based training toward models capable of handling entire 3D volumes. While progress has been made, especially in segmentation and volumetric representation learning, the field still lacks large-scale, general-purpose 3D foundation models comparable to their 2D counterparts.


\begin{table*}
\caption{Foundation Models specialized for CT and/or MRI. Type refers to whether the FM is a vision model (VM) or a vision-language model (VLM). Backbone and SSL objective show how the model was pretrained. Imaging domain shows which medical image types were used to pre-train the model. Dataset size for pre-training is given in number of 3D volumes, unless stated otherwise. Availability shows whether the weights and pretraining data are available online.}
\label{table:rad_3d}
\footnotesize
\begin{minipage}{\textwidth} \begin{center}
\centerline{\begin{tabular}{@{}llllllll@{}}
\toprule
\multirow{2}{*}{\textbf{Model}} & \multirow{2}{*}{Type} & \multirow{2}{*}{Imaging} & \multirow{2}{*}{Backbone} & \multirow{2}{*}{SSL Objective} & \multirow{2}{*}{Data Size} & \multicolumn{2}{c}{Availability} \\
\cmidrule(lr){7-8}
& & & & & & Weights & Data \\
\midrule
\textbf{DeSD}            & VM & CT  & 3D ResNet50                         & DeSD                  & 11K     & \xmark & \cmark \\
\textbf{SMIT}            & VM & CT, MRI & SWIN-small                        & MIM + Self-Distillation & 3643  & \cmark & \xmark \\
\textbf{Medical Transformer} & VM & MRI & ResNet-18 + Transformer           & MAE                    & 1783   & \xmark & \cmark \\
\textbf{M3FM}            & VLM & CT  & 3D CT-ViT + Transformer             & ITC                    & 163K   & \cmark & \cmark \\
\textbf{CLIP-LUNG}       & VLM & CT  & ViT-B/16 + ResNet18                 & ITC                    & 1010   & \xmark & \cmark \\
\textbf{Niu et al.}      & VM & CT  & 3D ViT                              & Region-Contrastive     & 684    & \xmark & \cmark \\
\textbf{MedBLIP}         & VLM & MRI & MedQFormer + BioMedLM\textsuperscript{a} & ITC             & 30K    & \cmark & \cmark \\
\textbf{MeTSK}           & VM & fMRI & STGCN                               & GraphCL                & 1415   & \xmark & \cmark \\
\textbf{Pai et al.}      & VM & CT  & 3D ResNet                           & Modified SimCLR        & 11.5K  & \cmark & \cmark \\
\textbf{M3D-LaMed}       & VLM & CT  & 3D ViT + LLaMA-2                     & ITC                    & 120K   & \cmark & \cmark \\
\textbf{Merlin}          & VLM & CT  & I3D ResNet152 + Clinical-Longformer  & Contrastive            & 25K    & \xmark\textsuperscript{a} & \xmark\textsuperscript{a} \\
\textbf{CT-CLIP}         & VLM & CT  & 3D CT-ViT + CXR-Bert                & ITC                    & 26K    & \cmark & \cmark \\
\bottomrule
\end{tabular}} \end{center}
\footnotesize  
(a) Planned to be released. 
(b) Authors experiment with multiple language encoders. Table shows their best performing model architecture.
\end{minipage}
\end{table*}

\subsubsection{Vision-Only Models}
While vision-language models have been predominant in radiology, driven by tasks like report generation and question answering, there is growing interest in developing vision-only foundation models. VFMs are better suited for spatially focused tasks such as detection and segmentation, which require precise understanding of the (3D) structure in medical imaging. 
For instance, \textit{Medical Transformer} \citep{jun2021medicaltransformeruniversalbrain} and research by \cite{niu_unsupervised_2022} employed multi-view and sequence-based approaches to handle 3D data, though these models only partially leveraged the full spatial potential of 3D volumes.

As the field progressed, various self-supervised learning techniques emerged. Initial approaches, such as those using contrastive learning, were inspired by vision-language models. Later, knowledge distillation techniques gained prominence, with models like \textit{DeSD} \citep{ye_desd_2022} and \textit{SMIT} \citep{jiang_self-supervised_2022} improving 3D segmentation through self-distillation and masked image modeling. The framework introduced by \textit{UniMiSS} \citep{xie2022unimiss} expanded on these concepts, accommodating both 2D and 3D data. Additionally, more recent models explored graph-based SSL methods, such as \textit{LVM-Med} \citep{nguyen_lvm-med_2023} and \textit{MeTSK} \citep{cui_meta_2023}, which use graph structures to enhance feature representation across different imaging modalities.

Some works also have started using the DINOv2 framework for medical image analysis tasks. For instance, \textit{RAD-DINO} \citep{perez-garcia_rad-dino_2024}, adapted the DINOv2 technique for medical imaging, showing strong performance in image registration and disease classification. Similarly, \cite{moutakanni2024advancing} apply DINOv2 to x-ray data. Their model, \textit{RayDINO}, is among the first VFMs developed specifically for X-ray imaging. Trained on 873,000 X-rays, RayDINO adopts a comprehensive approach by not relying on assumptions about predefined classes. Instead, it focuses on extracting general features from imaging data, enabling it for example to detect rare diseases that may not be well-represented in training data. Compared to CheXzero, KAD and UniChest, which rely on textual supervision, RayDINO surpasses them in tasks such as multi-label disease detection, segmentation, and even report generation, while also generalizing well to unseen populations, achieving these results solely through large-scale self-supervised learning on imaging data. 

Further advancements include models like \textit{VISION-MAE} \citep{liu2024visionmae}, which applied masked autoencoders to 3D data but was limited by its slice-level processing. \cite{pai_foundation_2024} introduced a cancer-focused model trained on 3D CT data, excelling in biomarker discovery despite its narrower focus. The \textit{Segment Anything Model} (SAM) has been tested across various studies, including \cite{yan2022sam}, \cite{mazurowski2023segment}, \cite{zhang2023customized}, and \cite{gao_desam_2023}, demonstrating its versatility in medical image segmentation while also facing challenges in cross-domain adaptation and efficient deployment.

\subsection{Concluding Remarks}

While radiology has not yet seen the same rapid development of foundation models as pathology, progress is being made, particularly with vision-language models. Early models in radiology focused mainly on tasks like disease classification and report generation using 2D images, often based on CLIP-like architectures or contrastive learning. Approaches like \textit{CheXAgent} highlight the value of increasing scale of training data. However, not all models have access to data on such large scale, and many models still rely on improving the quality of available data, using methods like knowledge enhancement and feature extraction.

In parallel, there has been growing interest in vision-only foundation models, particularly for tasks requiring detailed spatial understanding, such as segmentation, localization, and disease detection. This trend is partly inspired by the success of vision-only approaches in pathology, though the differences between these modalities mean it remains unclear whether similar results can be achieved in radiology. Vision-only models offer advantages in avoiding the challenges of creating high-quality paired vision-language datasets, such as the limited captions in PubMed datasets or the scarcity of large-scale radiology reports. However, they face their own hurdles, including the availability of large-scale 3D datasets and the computational resources required to train 3D architectures. Many models still rely on 2D slices or pretrained networks, and fully integrated 3D models are not yet widespread. Meanwhile, vision-language models remain indispensable for tasks that require integrating imaging data with clinical context and language understanding, such as report generation and question answering, though they continue to grapple with the limitations of existing datasets. Moving forward, the field may benefit from a balanced exploration of both paradigms, leveraging their respective strengths to address the diverse challenges in medical imaging.

Overall, foundation models in radiology are still in the early stages of development but have shown promising progress. Recent advancements emphasize the importance of scaling training data, improving the quality of available data, and shifting the focus to vision-specific tasks like segmentation and disease detection. Despite challenges such as the limited availability of large-scale 3D datasets and the computational demands of training 3D models, ongoing research in these areas is likely to lead to significant breakthroughs and enhanced integration of these models into clinical workflows.

\section{Ophthalmology}\label{section:ophthalmology}

The field of ophthalmology, which focuses on diagnosing and treating eye disorders, utilizes multiple specialized imaging modalities focused on the eye. Not only are these images used to diagnose common conditions such as diabetic retinopathy, glaucoma, and age-related macular degeneration (AMD), they can also reveal indicators of systemic conditions like early-onset Parkinson's disease and cardiovascular disorders. 

Ophthalmic images are particularly suited for AI applications for several reasons. One of the main factors driving the use of AI in this field is the availability of large datasets. Retinal imaging techniques like Color Fundus Photography (CFP) and Optical Coherence Tomography (OCT) are relatively non-invasive and less expensive compared to biopsies or radiological scans, which makes it easier to collect extensive datasets essential for training robust AI models.

\subsection{Foundation Models for Ophthalmology}

\subsubsection{Foundation Models for Fundus Photography and OCT}
The first foundation model for ophthalmology was RETFound by \cite{zhou_foundation_2023}. RETFound has a ViT model as its backbone, which is pre-trained on a large-scale dataset of over a million unlabelled retinal images using MAE during self-supervised learning. Two models are pretrained, either on OCT data only or CFP data only. After pre-training, these RETFound models are fine-tuned for specific disease detection tasks using explicit labels.
The models can identify and quantify diseases such as diabetic retinopathy and AMD. Furthermore, with minimal fine-tuning, they could also detect other conditions, such as heart failure, stroke, and Parkinson's disease. 

Expanding on RETFound, \cite{engelmann2024training} propose RETFound-Green, a variant of the original RETFound CFP model that focuses on efficiency. The model utilizing a novel token reconstruction pre-training objective, that helps it learn to identify features that are important for disease detection. This method enabled training with significantly less data and compute resources (as seen in Table \ref{table:opthalmology}) while maintaining comparable performance.
Furthermore, DERETFound by \cite{yan2024expertise} enhances performance by integrating stable diffusion techniques to generate synthetic data and incorporating medical expertise through text-tagging of images. This approach achieved comparable or superior results to RETFound using a fraction of the original dataset.
%
\begin{table*}[t]
\caption{FMs for ophthalmology. Type refers to whether the FM is a vision model (VM) or a vision-language model (VLM). Modality refers to the type of ophthalmological data the model pre-trained on. Backbone and SSL objective show how the model was pretrained. Data size shows the number of 2D images used for pre-training, while model size denotes the number of parameters used for the model. Availability shows whether the weights and pretraining data are available online.}
\label{table:opthalmology}
\footnotesize
\begin{minipage}{\textwidth}
\begin{center}  
\begin{tabular}{@{}lllllllllll@{}}
\toprule
\multirow{2}{*}{\textbf{Model}} & \multirow{2}{*}{Type} & \multirow{2}{*}{Imaging} & \multirow{2}{*}{Backbone} & \multirow{2}{*}{SSL Objective} & \multicolumn{2}{c}{Size} & \multicolumn{2}{c}{Availability} \\
\cmidrule(lr){6-7} \cmidrule(lr){8-9}
                       &                       &                           &                           &                                          & Data  & Model                  & Weights  & Data     \\
\midrule
\textbf{RETFound}       & VM  & CFP, OCT                  & ViT-Large                 & MAE                    & 1.6M  & 307M  & \cmark    & \cmark/\xmark\textsuperscript{a} \\
\textbf{FLAIR}          & VLM & CFP                       & ResNet-50                 & CLIP                   & 285K  & 23M   & \cmark    & \cmark \\
\textbf{VisionFM}\textsuperscript{g} & VM & Various\textsuperscript{c} & - & - & 3.4M & - & \xmark & \cmark/\xmark\textsuperscript{b} \\
\textbf{DRStageNet}     & VM  & CFP                       & ViT-Base                  & DINOv2                 & 93.5K & 86.9M & \xmark & \cmark \\
\textbf{DERETFound}     & VLM & CFP                       & ViT-Large                 & MAE                    & 150K + 1M\textsuperscript{d} & 307M & \cmark & \cmark \\
\textbf{RETFound-Green} & VM  & CFP                       & ViT-Small                 & Token Reconstruction   & 75K   & 22.2M & \cmark\textsuperscript{f} & \cmark \\
\textbf{EyeFound}       & VM  & Various\textsuperscript{e} & ViT-Large               & MAE                    & 2.8M  & 307M  & \xmark & \xmark \\
\bottomrule
\end{tabular} \end{center}

\footnotesize  
(a) All datasets are publicly available except for the MIDAS dataset, which is subject to controlled access via an application process. 
(b) All datasets are publicly available except for one private MRI dataset. 
(c) CFP, FFA, OCTA, OCT, Slit-Lamp, B-Scan Ultrasound. 
(d) 150K real and 1M synthetic generated images. 
(e) CFP, FFA, ICGA, FAF, RetCam, Ocular Ultrasound, OCT, Slit-Lamp, External Eye Photo, Specular Microscope, Corneal Topography. 
(f) Planned to be released. 
(g) No architectural details available.

\end{minipage}
\end{table*}
Another FM for ophthalmology that integrates textual expert-knowledge is FLAIR \citep{silvarodriguez2023foundation}, which uses CLIP to align image features from fundus photography with textual descriptions. This integration of expert knowledge has shown to enhance performance in diabetic retinopathy grading, glaucoma detection, and multiclass disease diagnosis \citep{silva2024importance}. Another model built for CFP data is DRStageNet \cite{men2023drstagenet}, which unlike the more generalist models described so far, is tailored specifically for diabetic retinopathy staging. DRStageNet employs a ViT backbone and DinoV2 for training. Its specialized architecture has demonstrated high accuracy and consistency in clinical settings. 

\subsubsection{Multi-Imaging Ophthalmology Foundation Models}
While previous models were exclusively trained on fundus photography or OCT, VisionFM \citep{qiu2023visionfm} integrates a variety of ophthalmic imaging modalities, including angiography, slit-lamp, and ultrasound. These imaging techniques can be used for visualizing other parts of the eye, not just retina, but also iris or cornea amongst others. 
VisionFM can diagnose diseases across these modalities using a single decoder, supporting a broad range of tasks like lesion segmentation and landmark detection, while generalizing efficiently to new imaging types. However, the paper does not yet provide specifics about the model architecture or training method, leaving the full scope of the results unclear at present.

More recently, EyeFound \cite{shi2024eyefound} was introduced as a general ophthalmic imaging FM, featuring a large ViT backbone and MAE pre-training, initialized with RETFound's weights. The model is trained on a dataset of 2.8 million images from diverse ophthalmic imaging modalities, as specified in Table \ref{table:opthalmology}. EyeFound achieves higher AUROC scores for glaucoma detection and DR evaluation, and it outperforms RETFound in multi-label disease classification and systemic disease prediction. Additionally, it is evaluated on a visual question-answering task and demonstrates faster convergence during fine-tuning. However, its performance is not universally superior; in some datasets, its diagnostic accuracy is comparable to RETFound, and in tasks like myocardial infarction prediction, it does not significantly surpass RETFound, indicating areas for improvement in handling diverse and imbalanced datasets.

\subsubsection{Techniques for Scaling Data}
Beyond these FMs, smaller-scale models have introduced novel methods to address the problem of data availability in retinal image analysis. 
\cite{alam_contrastive_2023} propose FundusNet, a Resnet50 model pre-trained through contrastive learning with Neural Style Transfer (NST) augmentations. NST enhances training data diversity by transferring image styles, which is then used in a contrastive learning framework to train the model to maximize the similarity of different views of the same image and minimize the similarity of different images. This approach allows FundusNet to achieve improved performance and generalization across different datasets.

Furthermore, 
\cite{zhang_annotation-efficient_2023} focus on annotation-efficient learning for OCT segmentation. Their method uses generative self-supervised learning to train a ViT model on OCT images, which is further fine-tuned with a CNN encoder head for dense pixel-wise predictions. To reduce the annotation workload, they develop a selective annotation algorithm based on the greedy approximation for the k-center problem, constructing a representative subset of the data for annotation and significantly reducing the annotation cost.

\subsubsection{Application of SAM for Ophthalmology}
Finally, some works focus on enhancing SAM for retinal image analysis. \cite{deng_sam-u_2023} propose SAM-U, a method that enhances the reliability of SAM in medical imaging through multi-box prompts and uncertainty estimation. It uses different prompts as a form of test-time augmentation to estimate the distribution of SAM predictions. Demonstrating their methods on fundus photography images, the authors show SAM's improved performance while also providing pixel-level uncertainty, making it particularly useful for medical image segmentation.
Another work that builds on SAM is SAM-OCTA \citep{wang_sam-octa_2023}, a method for local segmentation in Optical Coherence Tomography Angiography (OCTA) images. It fine-tunes a pre-trained SAM model using LoRA and utilizes prompt points for local segmentation of retinal vessels, arteries, and veins. It has shown state-of-the-art performance in common OCTA segmentation tasks, particularly in local vessel segmentation and artery-vein segmentation, which were not well-solved in previous works.

\subsubsection{Concluding Remarks}

Starting with RETFound, the field has seen significant improvements in efficiency and handling multiple imaging modalities. Models like RETFound-Green and DERETFound focus on reducing data and computational requirements without sacrificing performance, and methods are being proposed to
address challenges surrounding limited training data. Meanwhile, newer models like VisionFM and EyeFound stand out for their ability to handle various imaging modalities and perform multiple tasks. These advancements enhance the potential for these models to be integrated into clinical practice.
It should be noted however, that there is still a focus on retinal imaging and less on other parts of the eye. Furthermore, almost exclusively 2D data is used

Compared to trends in pathology and radiology, there is a predominance of vision models over vision-language models in ophthalmology. While VLMs like FLAIR and DERETFound incorporate expert knowledge, there are no foundation models that extensively utilize report data, as seen in radiology. This distinction likely stems from differences in clinical workflows and the role of textual data in decision-making.
Radiological images typically capture large anatomical structures, meaning lesions, fractures, or other abnormalities must be identified within a broader spatial context. Reports play an important role in contextualizing these findings, and incorporate for example spatial relationships, comparisons with prior scans, and clinical history \citep{pesapane2023advancements}. In contrast, ophthalmology diagnoses are more driven by direct image interpretation, with less reliance on paired textual descriptions \citep{lee2021recommendations, shweikh2023growing}. As a result, ophthalmology has fewer large-scale image-text datasets, limiting the development of vision-language foundation models in this domain.

Another challenge in the field is the lack of inter-model comparisons. Except for the derivatives of RETFound, later models seldom reference or build upon each other, often being evaluated against generic models trained on ImageNet rather than against each other. This practice limits the ability to effectively evaluate and compare different models. 

Overall, foundation models for ophthalmology are still in the early stages of development and face several challenges. Nonetheless, significant progress has been made, with recent efforts emphasizing efficiency and expanding the range of imaging modalities showing promise.
Continued research in these directions will likely lead to significant advancements and better integration of these models into clinical practice.

\section{Challenges and Future Directions}\label{section:challenges}

Foundation models have shown remarkable promise in medical image analysis, with improved generalization and robustness compared to traditional deep learning methods \citep{nguyen2023distribution, azizi_robust_2023}. However, several challenges need to be addressed to ensure their safe and effective deployment in clinical settings. These challenges fall into two broad categories: (1) technical barriers in model development, and (2) practical challenges in clinical integration. In the sections below, we outline each of these areas and offer directions for future research.

\subsection{Challenges in FM Technical Development}
The development of foundation models for medical imaging faces several technical challenges. This involves issues like limited access to large-scale datasets, high computational demands, and possible scaling limitations.

\subsubsection{Lack of Open-Source Large-Scale Clinical Datasets}
Unlike natural image datasets, medical imaging datasets often suffer from restricted access due to privacy regulations and/or institutional policies. As seen in this Tables 1-7, the availability of datasets varies quite a lot. In fact, many of the larger datasets fall under one of the following trends: \textit{1.} large datasets that were collected from public sources, such as Twitter or PubMed, and are therefore also made open-source, or  \textit{2.} large datasets that are proprietary, such as those of Virchow, Phikon, GigaPath or VisionMAE.
In addition, datasets collected from the internet typically do not include full volumes for 3D imaging domains, nor different magnification levels as would be present in WSIs. This means that there still exists a lack of large, clinical datasets that are publicly available, which can limit reproducibility and collaboration. Research into directions such as federated learning and differential privacy can be beneficial for solving these challenges \citep{li2025open, wang2024fedmeki}.

\subsubsection{Increased Computational Demands of Volumetric Data}
For many of the works in this review, it could be observed that larger datasets improved model performance, and that there is a trend towards scaling up. However, scaling is particularly challenging for computationally demanding imaging modalities, such as high-resolution WSIs, and volumetric CT or MRI scans, even outside the medical imaging domain \citep{zuo2024towards}. As foundation models scale up in both data and architecture, ongoing research should prioritize developing more efficient solutions specifically tailored for processing volumetric data, as seen for example in \cite{dominic_improving_2023}

\subsubsection{Pathology FMs Show Plateau in Scaling-Laws}\label{path_scale}

While the challenges in scaling up data and computational resources are still ongoing, pathology FMs have seen many recent successes by scaling up, downstream task performance generally improving with increased data and model size. However, some recent studies show a performance plateau when scaling along these axes \citep{ciga2022self, filiot2023scaling, aben2024towards, campanella2024clinical, chen2024towards, vorontsov2024foundation}, which can be attributed to several factors.

\paragraph{Data Quality over Quantity}
First, simply increasing data size may not be enough; better curation is needed to create a more diverse dataset that covers a range of organs, pathologies, and staining types. For example, \cite{dippel2024rudolfv} demonstrated that Virchow’s performance could be matched with a curated dataset containing ten times fewer slides. Furthermore, Atlas \citep{alber2025novel} could even outperform Virchow despite having a much smaller pre-training dataset size, by focusing on including a much broader range of scanners, stainings and magnifications. 
Similarly, \cite{filiot2024phikon} demonstrated that a smaller ViT-B/16 model pre-trained on 4 million patches using iBOT outperformed Virchow2, a model pre-trained on 350 times more histology images, across three external validation cohorts. This underscores the importance of targeted training strategies over brute-force scaling.

\paragraph{Lack of Diverse Benchmarks}
Furthermore, downstream task comparisons often rely on public challenge datasets, but a broader range of benchmark tasks is needed to fully assess pathology FMs. \cite{zimmermann2024virchow} found performance on public tile-level benchmarks plateauing, and \cite{campanella2024clinical} noted more variability in performance when using challenging biomarker prediction tasks as benchmarks.
To address these challenges, frameworks such as \texttt{Patho-Bench} \citep{zhang2025accelerating} and \texttt{eva} \citep{gatopoulos2024eva} provide standardized evaluation tools, enabling systematic benchmarking across diverse datasets, tasks, and evaluation protocols.

\paragraph{Adapting to the Medical Domain}
Finally, current SSL frameworks are optimized for natural images, not pathology images, which have distinct characteristics such as lack of canonical orientation and color variation, as well as interpretation differences depending on the field of view \citep{kang2023benchmarking}. SSL that use for example data augmentations or pretext tasks, such as colorization or rotation prediction frameworks, may need domain-specific adaptations for medical imaging data. 
\cite{filiot2024phikon} further showed that, while scaling general-purpose FMs like GigaPath, H-Optimus-0, Phikon-v2, UNI, and Virchow2 often benefits downstream tasks, these models do not excel uniformly across all organs or tasks.

Although studies have explored domain-specific adjustments (e.g., weight initialization, data augmentations, magnification levels), both in pathology \citep{wang2022transformer, hua2023pathoduet, kang2023benchmarking, wu2023brow, aben2024towards, ma2024towards, jaume2024transcriptomics, jaume2024multistain, roth2024low, yun2024enhancing, zimmermann2024virchow} and other medical imaging domains \citep{ niu_unsupervised_2022, ye_desd_2022, jiang_self-supervised_2022, wu_towards_2023}, many open questions remain about developing successful SSL algorithms for medical imaging. As foundation models for all medical imaging domains are increasingly being developed, these questions become relevant for ensuring scalability, generalizability, and the effective utilization of resources across all medical imaging domains.

\subsection{Challenges in Applying FMs to Clinical Settings}
The transition from research to clinical practice for medical imaging foundation models involves overcoming several key challenges. These include ensuring the models’ interpretability, addressing potential biases, maintaining fairness, and navigating regulatory requirements.

\subsubsection{Explainability}

Foundation models for medical image analysis will need to provide transparent and interpretable results, enabling clinicians and researchers to understand how the model is making predictions and what features are driving those predictions.
However, the complexity of foundation models makes it difficult to understand how they make predictions, identify biases, and explain errors, which can have serious consequences in healthcare \citep{marcus2024artificial}.
Explainable AI techniques can help address this issue. For example, \cite{abbas_xdecompo_2022} introduced an interpretable decomposition approach for tumor classification, while \cite{pham_i-ai_2023} developed a controllable and interpretable system for chest X-ray diagnoses. \cite{deng_sam-u_2023} add pixel-wise uncertainty estimation to SAM, making it possible to evaluate the the reliability of segmented lesions. 

\subsubsection{Robustness and Domain Generalization}

Foundation models in medical imaging must remain robust to variations in data acquisition, staining, scanning equipment, and institutional protocols. Without this robustness, models risk producing unreliable or misleading results in real-world deployment. For example, \cite{de2025current} found that many current pathology FMs encode confounding factors such as medical center-specific artifacts more strongly than relevant biological signals. They proposed the robustness index, a quantitative metric to evaluate how well model representations are driven by biological rather than spurious features.

Similarly, \cite{filiot2025distilling} demonstrated that a distilled pathology FM (H0-mini) outperformed larger models on robustness metrics, showing resilience to staining and scanning variability on the PLISM dataset. This highlights that model size does not always correlate with robustness, and that distillation or other targeted training techniques may help improve domain generalization.

Addressing robustness is crucial for clinical deployment, particularly when models are transferred across institutions or used in out-of-distribution settings.

\subsubsection{Fairness, Bias and Robustness}

Bias presents a significant challenge for FMs, particularly in clinical contexts where the accuracy and fairness of predictions are critical. Models trained on biased datasets risk producing biased predictions, which can disproportionately affect underrepresented populations \citep{glocker_risk_2023}. Addressing this requires diverse and representative training data, alongside robust fairness evaluation frameworks. Studies such as \cite{czum2023bias} and \cite{khan_how_2023} highlight the importance of identifying and mitigating biases in FMs. Specific approaches that aim to reduce bias include cross-lingual pre-training, as proposed by \cite{wan_med-unic_2023}, which helps reduce bias in vision-language models. These solutions are vital to ensure FMs provide equitable benefits across diverse patient populations.

Furthermore, ensuring the robustness of foundation models to variables like scanner type, staining protocols, and institutional differences is crucial for their generalizability across clinical settings. \cite{de2025current} found that many current pathology foundation models encode confounding factors, such as medical center-specific staining artifacts, more strongly than relevant biological signals like tissue or cancer type. To address this, they proposed the robustness index, a quantitative metric that measures the extent to which model representations are driven by biological features rather than spurious confounders, providing a practical tool for evaluating model reliability.
\cite{filiot2025distilling} explore the distillation of a large foundation model into the much smaller \textit{H0-mini}. Their model, demonstrates excellent robustness to variations in staining and scanning conditions, significantly outperforming other current pathology FMs. 

Finally, with the growing trend of scaling up foundation models, there is a risk of making them inaccessible to institutions with limited computational resources. It should therefore be encouraged to develop more efficient FM architectures to improve accessibility \citep{engelmann2024training, pohjonen2024histoencoderdigitalpathologyfoundation}. Additionally, parameter-efficient fine-tuning strategies, as highlighted by \cite{dutt_parameter-efficient_2023}, offer a promising pathway for enabling widespread adoption while minimizing computational demands.

\subsubsection{Regulation}

For FMs to be successfully adopted in clinical practice, their risks must be addressed through comprehensive regulation. Medical imaging FMs, like other AI systems, can produce unreliable outputs, such as hallucinations; plausible but incorrect predictions that could lead to misdiagnoses or inappropriate treatments. Recent studies propose methods for detecting hallucinations in medical vision-language FMs \citep{sambara2024radflag, yan2024med}, but more research is needed to extend these approaches to vision-based FMs.

Continual learning is a promising research direction to ensure that FMs can adapt to new tasks, disease variants, or data without forgetting previously learned knowledge \citep{wang2024comprehensive, yi2023towards}. This adaptability is crucial in the rapidly evolving medical field. 
\newpage
\bibliographystyle{unsrtnat}
\bibliography{references}
\end{document}